\documentclass[preprints,article,accept,pdftex,moreauthors]{Definitions/mdpi}

\firstpage{1} 
\pubvolume{1}
\issuenum{1}
\articlenumber{0}
\pubyear{2026}
\copyrightyear{2026}
\datereceived{ } 
\daterevised{ } %
\dateaccepted{ } 
\datepublished{ } 

\usepackage[svgnames]{xcolor}
\usepackage{braket}
\usepackage{dcolumn}
\newcolumntype{d}[1]{D{.}{.}{#1}}
\graphicspath{{figs/}} %

\Title{Electrons Hopping across a Molecular Network: Spectra and Symmetries}

\Author{Ludwig Schulz\orcidA{}, Max Best\orcidC{}, and Carsten Henkel*\orcidB{}}

\AuthorNames{Ludwig Schulz, Max Best, and Carsten Henkel}

\address[1]{%
University of Potsdam, Institute of Physics and Astronomy, Germany; carsten.henkel@uni-potsdam.de
}

\abstract{
We investigate interacting spinless electrons on finite molecular \deletethis{ring} networks described by a tight-binding Hubbard Hamiltonian. 
The interplay between \deletethis{lattice} \replaceby{network} geometry, Coulomb interactions and discrete symmetries is analysed for rings with 
$L=3,4,5,6$ nodes, filled with one, two or three electrons.
Special attention is devoted to the role of the network symmetries in determining the structure of the many-body spectrum and the Mulliken classification of the eigenstates. 
Using group-theoretical methods, we examine the evolution of the spectra in the presence of an external magnetic flux.
The Zeeman effect lifts degeneracies and results in combination with the Coulomb interaction to avoided crossings in symmetry sectors. 
We identify a qualitatively distinction between systems with an even and odd number of particles. 
At half-filling, particle-hole symmetry (duality) protects selected symmetry sectors against Zeeman splitting.
\\[2ex]
\emph{Physics} \textbf{8} (2026) 73 [\href{https://DOI.org/10.3390/physics8040073}{DOI:10.3390/physics8040073}],
\\
\href{https://www.mdpi.com/journal/physics/special_issues/Quantum100}{Special Issue: Quantum Theory 100 Years Later: Advances on Foundations and Applications}
}


\newcommand{\hop}{t}   %
\newcommand{\gauge}{Q} %
\usepackage[normalem]{ulem}
\newcommand{\deletethis}[1]{\ignorespaces}
\newcommand{\replaceby}[1]{#1}
\newcommand{\addtext}[1]{#1}
\begin{document}
\section{Introduction}
Correlated electrons on discrete lattices represent a fundamental model system in many-body physics. In particular, the graphs of molecular tight-binding models provide a natural framework for studying the interplay between geometry, symmetry, and interaction effects.
\deletethis{While} These concepts are being put to work in \emph{ab initio} methods like tight-binding density functional theory \cite{Spiegelman_2020},
\addtext{with the aim of reproducing electronic and vibrational properties of actual molecules. 
In this paper the focus is different: we revisit the interplay between interactions and statistics in many-electron states. 
For free electrons in a metal, this gives rise to exchange-correlation energies, as initiated by Wigner's seminal paper \cite{Wigner1934}. 
The molecular networks we consider here are modeled with a tight-binding Hubbard Hamiltonian \cite{Hubbard1963}.
This description reduces a molecule that is hosting a moderate number of electrons, to a few essential features: the nuclei provide localized orbitals or vertices on a graph, and the molecular bonds (links, edges) allow for hopping between orbitals. 
The long-range Coulomb interaction is included, while on-site double occupancy is forbidden by the Pauli principle, assuming for simplicity that all electronic spins are aligned.
The Zeeman splitting in a magnetic field appears when the phase of the hopping matrix element is modified according to the Peierls substitution \cite{Peierls1933}.
Our model contains the main ingredients to generate strongly correlated many-body states, and we indeed observe interaction signatures beyond the independent-particle approximation. 
The discrete distances between the network sites impede, however, to study in its cleanest form the concomitant phenomenon of Wigner crystallization%
\footnote{\sethlcolor{Azure}\hl{%
A Wigner crystal is a system, typically with few electrons only, where the average ground-state density is periodic. This is supposed to arise from the Coulomb interaction and shows a different peak spacing compared to molecular orbitals in the form of standing waves. The spacing is also unrelated to the discrete lattice underlying the models studied here. Actually, the networks in this paper are too small to display Wigner crystallisation: two electrons occupying the graphene flake sketched in Fig.\:D1 would correspond to a Wigner-Seitz radius $r_s \approx 10$, while Wigner crystals are expected to occur for $r_s > 30{-}100$ depending on the dimension (2 or 3).
}} 
that appears in low-dimensional systems \cite{Wigner_molecules}.
Instead, the finite size and specific geometry of the network give rise to a rich set of discrete symmetries, as illustrated in Fig.\:\ref{fig:sketches}, left, which provide the natural language for classifying the resulting many-body spectrum.%
}
We find in particular that the network displays qualitative  (``topological'') differences for even or odd particle numbers.
\deletethis{we focus in this  work on a few essential features and investigate interacting spinless electrons on molecular networks within the framework of a Hubbard Hamiltonian \cite{Hubbard1963}. 
The network graph contains nodes representing localized orbitals, and its edges correspond to tunneling (hopping) between sites. 
In addition, we include the long-range Coulomb interaction, while on-site double occupancy is forbidden by the Pauli principle.
The system spectra are particularly rich when the Zeeman splitting in a magnetic field is allowed for \cite{Peierls1933}.
The formulation is based on second quantization using fermionic creation and annihilation operators. 
Particular attention is paid to discrete symmetries specific to the finite size of the network,
as illustrated in Fig.\:\ref{fig:sketches}, left.
They influence the structure of the Hamiltonian, spectral degeneracies, and the classification of eigenstates.
}

The structure of the paper is as follows. 
In Sec.\:\ref{sec:observables} we introduce the tight-binding Hubbard Hamiltonian for spinless fermions on ring networks.
Particular emphasis is placed on the Fermi sign appearing in the many-body hopping matrix and on the gauge transformation that moves the impact of the magnetic field to a single link (that closes the ring).
Relevant observables next to the energy are \deletethis{in particular} the persistent current $I$ and the magnetisation $M$.

In Sec.\:\ref{sec:symmetries} we discuss the discrete symmetries of the model. 
We introduce the point groups \deletethis{symmetries} of ring networks (cyclic and dihedral group) and explain the classification of many-body states into irreducible representations. 
The general construction is illustrated for the triangle with one and two \deletethis{spinless} electrons,
after which is introduced the particle-hole transformation (duality) and its action on the hopping and Coulomb terms of the Hamiltonian. 

In the core section Sec.\:\ref{sec:results} of the paper, the symmetry-based classification is applied to flux-dependent spectra of rings with $L=4,5,6$ sites. 
In these examples, the Zeeman effect is manifest, the Coulomb interactions remove degeneracies among irreducible representations (avoided crossings), and particle-hole duality explains ``protected'' degeneracies.
In Sec.\:\ref{sec:magnetisation-and-rotating-flow}, we complement the energy spectra by a representation in the energy-magnetisation plane, that provides a transparent picture for the flux dependence of the persistent currents.

\section{Model and Methods}

\subsection{Basic Observables}
\label{sec:observables}

\subsubsection{Hubbard Hamiltonian}
\label{sec:Hubbard-Hamiltonian}

\begin{figure}
    \centering
    \raisebox{-0.5\height}{%
            \includegraphics[width=0.35\linewidth]{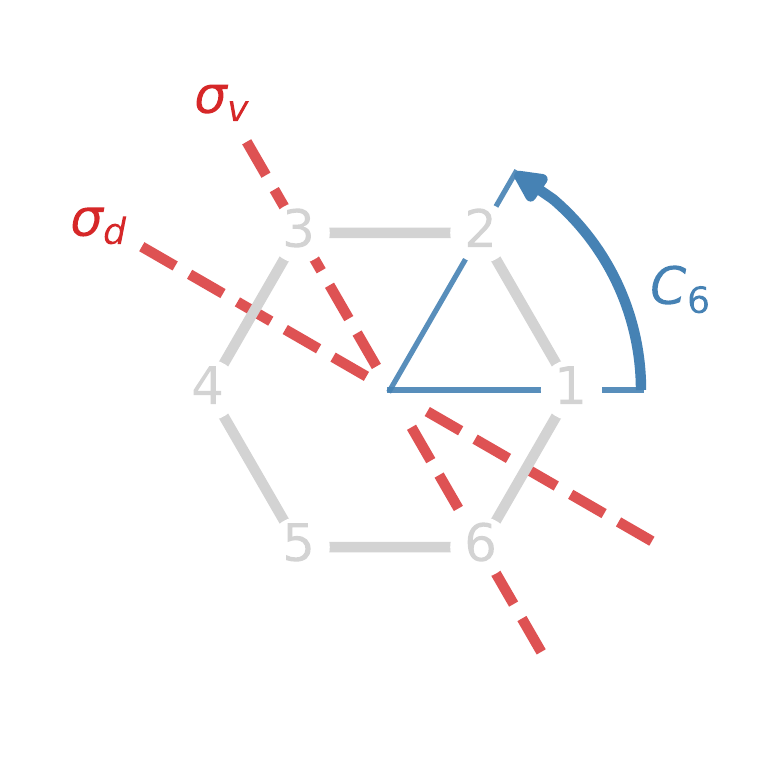}
    }
    \raisebox{-0.5\height}{%
        \includegraphics[width=0.35\linewidth]{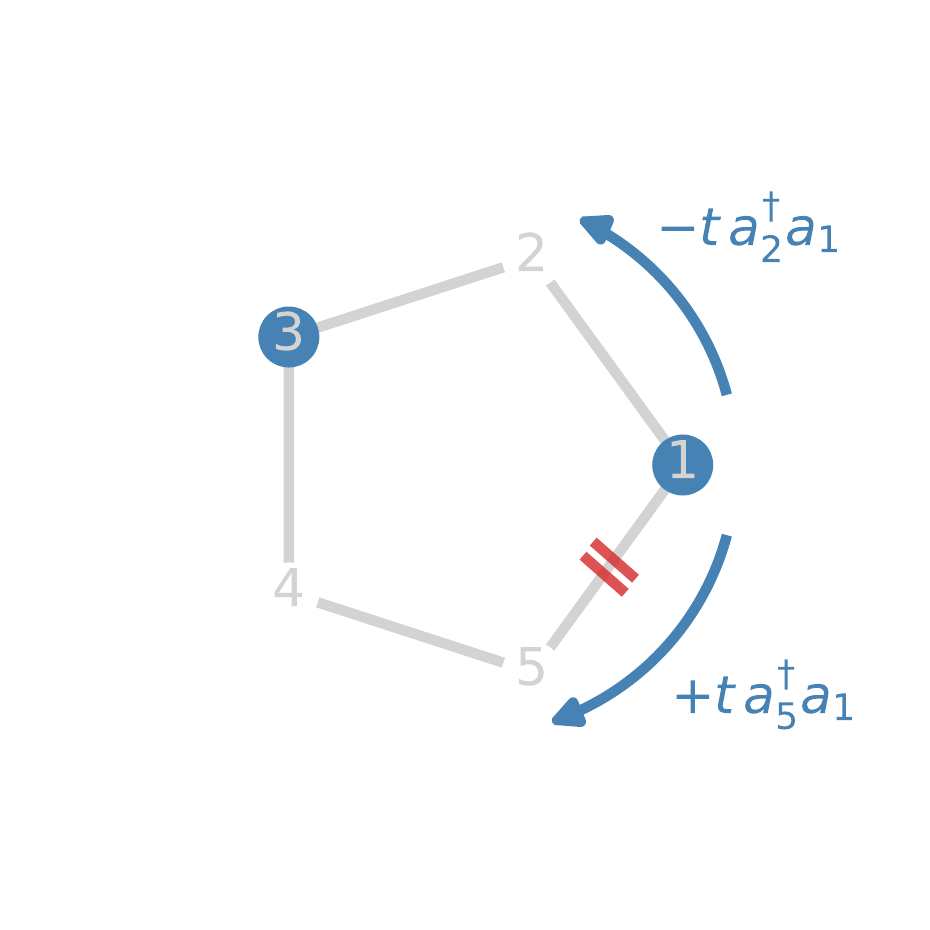}%
    }

    \caption[]{Typical features of ring networks. 
    Left: mirror axes and elementary rotation for an even number of sites.
    \addtext{For rings with an odd number of sites, the $\sigma_d$ mirror axis is absent.}
    Right: sign of hopping matrix elements. 
    The amplitudes for an electron hopping from site $1$ to neighboring sites may have opposite signs.
    After the hop $1\rightarrow5$ across the ``ring-closing'' edge (red barrier), a reordering is needed to restore the canonical ordering of creation operators.
    This leads to a minus sign if there is an odd number of other electrons on the ring.}
    \label{fig:sketches}
\end{figure}

The system is described by a tight-binding Hamiltonian for interacting spinless Fermions \cite{Hubbard1963}
\begin{equation}
\label{eq:ExtendedHubbard}
H =
- \sum_{s \ne t}
\left(
\hop_{st}\,
a_t^\dagger a_s^{\phantom\dag}
+
\mathrm{h.c.}
\right)
+
\sum_{s}
U_{s} n_s
+
\frac12 \sum_{s \ne t}
V_{st} n_s n_t 
\,.
\end{equation}
Here $\hop_{st}$ denotes the hopping amplitude from the source site $s$ to the target site $t$, 
$U_s$ is the on-site potential, 
and $V_{st}$ the two-body interaction, 
with $n_s = a_s^\dagger a_s^{\phantom\dag}$ being the local occupation number operator. 
We typically take the long-range Coulomb potential $V_{st} = e^2/4\pi\varepsilon_0 |\mathbf r_s - \mathbf r_t|$ with the site positions $\{ \mathbf r_s \}$.
The total particle number operator is $N = \sum_s n_s$,
\addtext{and all electrons are in the same spin state.}
The geometry underlying the model is represented by a graph, see Fig.\:\ref{fig:sketches}: nodes correspond to localized sites (one orbital per site), while edges (links) describe tunneling between neighboring sites. 
The connectivity of the network is encoded in the adjacency matrix. 
It has the same pattern of nonzero elements as the hopping matrix $\mathbf \hop = (\hop_{st})$.
Indeed, the adjacency matrix can be used to define a discrete Laplace operator on the network graph.

The many-body basis is constructed in Fock space as
$|n_1,n_2,\dots,n_L\rangle$, with occupation numbers $n_s = 0,1$ (Pauli principle). 
For a system with $L$ lattice sites and $N$ particles, the fixed-particle-number Hilbert space $\mathcal H_N$ has dimension
\begin{equation}
    \label{eq:dimension-HN}
\dim \mathcal H_N = \binom{L}{N}.
\end{equation}
Summing over all $N = 0, \ldots, L$, we get $2^L$ states in total. 
They can be conveniently enumerated using $n_L,\dots,n_2,n_1$ as binary digits.
Since the fermionic operators anti-commute, the matrix elements of the hopping operators $a_t^\dag a_s^{\phantom\dag}$ have signs that depend on the particle number,
as sketched in Fig.\:\ref{fig:sketches}, right.
The sign depends on the choice of an ordered list of local creation operators to generate Fock-space basis vectors (see Appendix\:\ref{a:OneToMany}).
If an electron has to hop from site $s$ to site $t$, the matrix element of $-\hop_{st} a_t^\dag a_s^{\phantom\dag}$ is $-\hop_{st}$ when the number of electrons occupying sites with \deletethis{numbers} \replaceby{labels} between $s$ and $t$ is even.
If this number is odd, the matrix element is flipped to $+\hop_{st}$.
This rule is rooted in the anti-commutation relations for fermionic operators and is justified in Sec.\,\ref{a:a-aDag-Fermi-sign}.
It only applies on the non-zero off-diagonal elements of the \addtext{graph} adjacency matrix \deletethis{of the hopping graph} and determines the representation $H_N$ of the \deletethis{hopping} Hamiltonian in a sector of fixed $N$.
The diagonal elements of $H_N$ in the occupation number representation (Fock basis) are determined by the interaction energy.

A magnetic field is incorporated via the Peierls substitution \cite{Peierls1933}
\begin{equation}
\hop_{st} \mapsto \hop_{st} \, {\rm e}^{ {\rm i} \theta_{s,t}}
\qquad \text{with} \quad
\theta_{s,t}
=
\frac{e}{\hbar}
\int_{\mathbf r_s}^{\mathbf r_t}
\mathbf A(\mathbf r)\cdot d\mathbf r ,
\end{equation}
where $\mathbf A(\mathbf r)$ denotes the vector potential and the integral is taken along the network link connecting $\mathbf r_s$ and $\mathbf r_t$. 
The complex hopping elements modify the phases of the electronic wave function and lead to flux-dependent spectra and persistent currents \cite{Lin2023}.

In this paper, we pay particular attention to the role of discrete symmetries and their influence on the many-body spectrum.
Mostly ring networks will be considered with periodic boundary conditions,
see Fig.\:\ref{fig:sketches}.
All sites being equivalent, we may take $U_s = 0$, and the Hamiltonian in Eq.\,~(\ref{eq:ExtendedHubbard}) becomes
\begin{equation}
\label{eq:SystemHamiltonian}
H
=
T + V
=
-\sum_{s=1}^{L}
\left(
\hop\,
\mathrm e^{\mathrm i\theta}
a_{s+1}^{\dagger}a_{s}
+
\text{h.c.}
\right)
+
\frac12 \sum_{s \ne t} V_{st} \hat n_s \hat n_{t}
\,,
\end{equation}
where $a_{L+1}\equiv a_1$. 
In a homogeneous magnetic field, the edges carry uniform Peierls phases $\theta = \frac{2\pi}{L}\frac{\Phi}{\Phi_0}$ with the magnetic flux $\Phi$ subtended by the network and $\Phi_0 = 2\pi\hbar/e$ denoting the flux quantum. 
An alternative form convenient for calculations can be achieved 
via a gauge (or Jordan-Wigner) transformation $\gauge$ that re-defines the phase of the $a_s$
\begin{equation}
    \label{eq:JordanWigner}
    \gauge\, a_s \, \gauge^{-1} = {\rm e}^{{\rm i}s \theta}\, a_s
    \,.
\end{equation}
By a calculation similar to the one of Appendix\:\ref{a:duality-and-hopping}, one finds the result that in 
$\gauge \, H \, \gauge^{-1}$, 
the Peierls phases vanish everywhere, except on the ring-closing edge between $s = L$ and $1$, where it is 
$\theta_{L,1} = L \theta = 2\pi\frac{\Phi}{\Phi_0}$. 
This illustrates that the dependence on the magnetic flux is periodic with the quantum $\Phi_0$, as is familiar, e.g., from the Aharonov-Bohm effect ~\cite{AharonovBohm1959}.

For the ring geometry, a Bloch \emph{Ansatz} $\psi_q(s) = L^{-1/2} \exp{(2\pi{\rm i} \, s q/L)}$ provides an explicit solution for the one-electron spectrum 
\begin{align}
    \label{eq:one-electron-spectrum}
    \varepsilon(q) = -2 \hop \cos \left[\frac{2 \pi}{L}\left( q - \frac{\Phi}{\Phi_0} \right)\right]
    \,.
\end{align}
This state is created by the Fourier-transformed operator
$c_q^\dag = \sum_s \psi_q(s) \, a_s^\dag\,$.
The periodic boundary condition $\psi_q(1) = \psi_q(L+1)$ fixes the Bloch momenta to be
$q = 0, \pm 1, \ldots$ 
until $|q| \le L/2$, which are defined only modulo $L$.
In particular for even $L$, the states $q = \pm L/2$ are the same.

\subsubsection{Current and Magnetisation}

Relevant observables of the system include the persistent current $I$ \cite{Meden2003}, the magnetization $M$, and the magnetic susceptibility $\chi$ \cite{London1937}.
They are defined through derivatives of the Hamiltonian $H$,
\begin{equation}
\label{eq:magnetisation-operator}
I
= -\frac{\partial  H}{\partial \Phi}
\,,\qquad
M
= -\frac{\partial  H}{\partial B} 
= A I
\,,
\end{equation}
where $B$ is the magnetic field and $A$ the ring area.
The current operator is explicitly
\begin{equation}
\label{eq:current-operator}
I = \frac{ {\rm i} \, e \hop }{ \hbar L }
\sum_s \left( {\rm e}^{{\rm i} \theta} a_{s+1}^\dag a_s^{\phantom\dag} - \text{h.c.} \right)
\,.
\end{equation}
Its expectation value in the Bloch state of Eq.\,(\ref{eq:one-electron-spectrum}) is
$\sim \sin \left(\frac{2 \pi}{L}\left[ q - \frac{\Phi}{\Phi_0} \right]\right)$.
An example how energy and magnetization change with the flux $\Phi$ can be found in Fig.\:\ref{fig:E-M-Flow} of Sec.\:\ref{sec:magnetisation-and-rotating-flow}.

\subsection{Symmetries}
\label{sec:symmetries}
\subsubsection{Point Group: Geometric Symmetry}
\label{sec:implement-point-group-symmetry}

The Hamiltonian is invariant under the geometric symmetries of the underlying graph, 
which form the discrete point group $G$. 
Examples for ring networks are the cyclic groups $C_L$ generated by rotations and the dihedral groups $D_L$ containing both rotations and reflections of the ring \addtext{(Fig.\:\ref{fig:sketches}, left)}.
A symmetry element is represented as a unitary transformation $g$ with the property $g H g^\dag = H$. 
This constrains the energy spectrum and enables us to decompose the Hilbert space into symmetry sectors labelled by irreducible representations (irreps) of $G$ \cite{Atkins, Tinkham}. 
Group representation theory implies that degenerate levels typically require a non-Abelian symmetry group (otherwise they are called ``accidental'').
When many-body states (configurations, Slater determinants) are built from tensor products of one-electron basis states, the reduction of product representations into irreps amplifies the understanding of the many-body energy spectrum. 
(The reduction procedure is analogous to the addition of angular momenta.)
This explains, for example, degeneracies or their lifting.

On a ring network, the generator $C$ of the cyclic group $C_L$ commutes with the Hamiltonian, and simultaneous eigenstates of both $H$ and $C$ can be found \addtext{that generalize the Bloch states of Eq.\,(\ref{eq:one-electron-spectrum}).}
The corresponding quantum number $q$ \addtext{(the total momentum)} is defined through the eigenvalue

\begin{equation}
\label{eq:good-q-numbers}
C\,\ket{\psi_q}
=
\mathrm e^{ - 2\pi \mathrm i q/L}
\ket{\psi_q},
\end{equation}
and labels the one-dimensional irreps of $C_L$. 
Usually, the condition $C^L = 1$ for the cyclic group holds so that the eigenvalues $q$ are integers and add modulo $L$.
This is the many-body equivalent to the periodic boundary conditions mentioned above for one-electron orbitals, and leads to the values $q = 0, \pm 1, \ldots$ until $|q| \le L/2$.
We shall see, however, that in certain situations, a twisted boundary condition arises and leads to half-integer quantum numbers (see Appendix~\ref{a:magnetised-dihedral-irrep}).

Note that each sub-space with $N$ electrons will have its own representation of a symmetry element (similar to the hopping signs discussed above).
The action on Fock space can be found from the representation~(\ref{eq:site-occupation-state}) and the convention
$g a^\dag_s g^{\dag} = a^\dag_{\pi(s)}$ where $\pi(s)$ is the image of the site $s$ under the geometric symmetry transformation behind $g$. 
There are signs that appear by re-ordering the creation operators $a^\dag_{\pi(s)}$ into the conventional order behind Eq.\,(\ref{eq:site-occupation-state}).
They are necessary to get an implementation of $g$ that commutes with the matrix representation $H_N$. \deletethis{in a fixed-$N$ subspace}

For many-body states, the total momentum is obtained by summing over occupied single-particle quantum numbers,
\begin{equation}
q
=
\sum_{\alpha=1}^{N}
q_\alpha
\pmod L.
\end{equation}
Consequently, the many-body Hilbert space can be decomposed into independent symmetry sectors labelled by $q$, which remain invariant under the action of the Hamiltonian.
This situation is typical for the cyclic group $C_L$ which is commutative.
In the absence of a magnetic field, however, the point group is the larger dihedral group $D_L$ and also contains mirror reflections $\sigma$ at planes perpendicular to the ring,
see Fig.\:\ref{fig:sketches}, left.
(One typically distinguishes for even $L$ between symmetry planes $\sigma_v$ going through a node and planes $\sigma_d$ going through two opposite edges. The two indeed belong to distinct conjugacy classes \cite{Atkins, Tinkham}.)
The generators $C$ and $\sigma$ of $D_L$ satisfy the fundamental relation
\begin{equation}
	\label{eq:dihedral-intertwining}
\sigma \, C = C^{-1} \, \sigma
\end{equation}
Equation~(\ref{eq:dihedral-intertwining}) illustrates that $D_L$ is not a commutative group, hence two-dimensional irreps exist and are given the Mulliken symbol $E_{q}$ with $q > 0$.
They can be spanned by states $|\psi_{\pm q} \rangle$ which are mapped one onto the other by a reflection.
[Apply Eq.\,(\ref{eq:dihedral-intertwining}) to $|\psi_{+q} \rangle$.]
The case $q = 0$ (and $q = \pm L/2$ for even $L$) corresponds to a one-dimensional representation with Mulliken label $A$ (and $B$), respectively.
Under dihedral symmetry, the \addtext{$A$, $B$} representations can be given an additional quantum number (subscript $1, 2$), describing their even/odd parity under $\sigma$:
the zero-momentum ground state for a single electron has the symmetry $A_1$, while $A_2$ describes, for example, a $q = 0$ singlet built from two electrons occupying momentum states $\pm 1$.
--
We sketch in Appendix~\ref{a:magnetised-dihedral-irrep} how the full dihedral symmetry can be implemented even with a nonzero magnetic flux, using anti-linear transformations.

\begin{figure}
    \centering
    \includegraphics[width=0.45\linewidth]{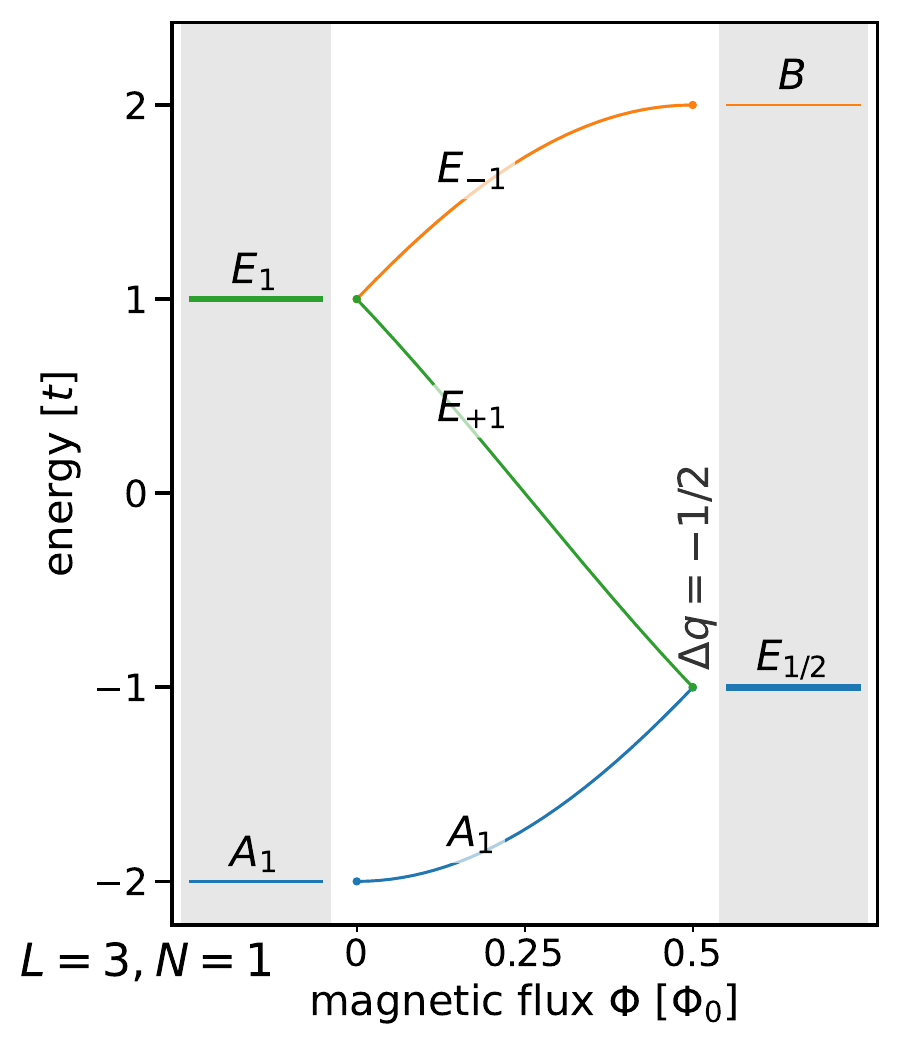}
    \hspace*{2em}
    \includegraphics[width=0.45\linewidth]{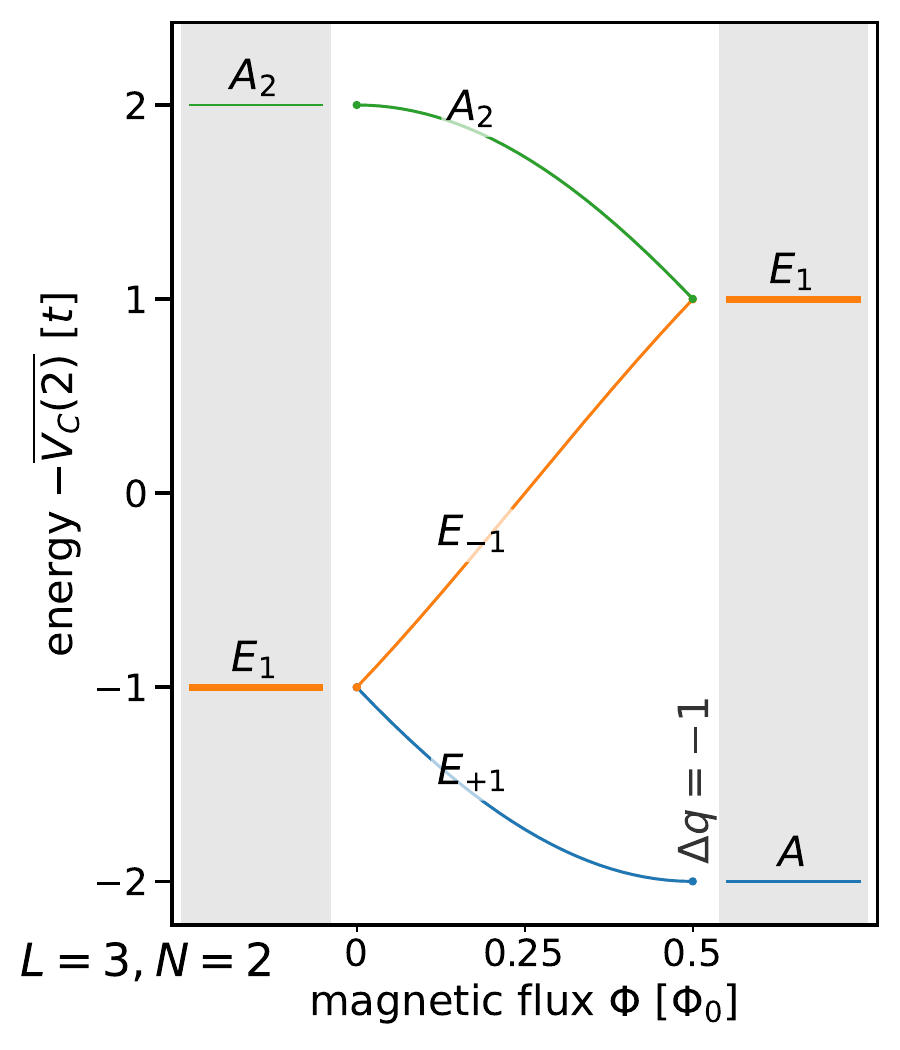}
    \caption[]{Zeeman spectrum for triangular ring with (left) one and (right) two electrons. 
    }
    \label{fig:spectrum_L3}
\end{figure}

\subsubsection[Example L = 3]{Example $L = 3$}

These concepts are illustrated in the Zeeman spectra for a triangular network with
$L = 3$ sites. 
Figure\:\ref{fig:spectrum_L3} shows the spectra calculated for $N = 1, 2$ electrons.
In this and the following plots, we subtract the mean Coulomb interaction $\overline{V_C(N)}$ for $N \ge 2$.
When comparing to non-interacting electrons (gray lines), only a differential Coulomb shift remains whose average is zero.

On the triangular network, the three number states $|1,1,0\rangle$, $|1,0,1\rangle$, $|0,1,1\rangle$ are degenerate with respect to interactions \addtext{because the two electrons always have the same distance $a$},
and we have $\overline{V_C(2)} = V(a)$ with the edge length $a$.
(See Appendix~\ref{a:Coulomb-scaling} for the scaling of $\overline{V_C(N)}$ with $N$ and network size $L$.)
--
The $N = 3$ case is not shown; 
it reduces to the fully occupied network $|1,1,1\rangle$ with pure Coulomb energy $3V(a)$.

The solid lines in the left plot are given by one-electron orbitals of Eq.\,(\ref{eq:one-electron-spectrum}).
They illustrate the linear Zeeman splitting of the momentum states in the $E_1$ term ($q = \pm 1$), having opposite persistent currents $I \sim \pm \sin\!\big[ \frac{2\pi}{3}(1 \mp \Phi/\Phi_0) \big]$.
The one-dimensional irreps $E_{\pm 1}$ with a signed subscript can be understood as a ``magnetised form'' of the dihedral symmetry and are discussed in Appendix~\ref{a:magnetised-dihedral-irrep}. 

The symmetries for two electrons (right plot) in zero field can be understood from the \emph{Aufbau} principle (see also Fig.~\ref{fig:E-M-Flow} bottom right):
the one-particle ground state $A_1$ has zero momentum, while the $N=2$ ground state is degenerate and corresponds to the \deletethis{two} configuration $a_1e_1$ 
(lowercase letters for single-particle orbitals) \addtext{with two states} of total momentum $q = \pm 1$.
The two electrons add their energies to form the $N=2$ ground state, interactions providing the shift $\overline{V_C(2)}$.
Conversely, the doubly occupied $e_1^2$ configuration forms the $A_2$ excited state.
This is the only antisymmetric configuration in the decomposition rule $E_1 \times E_1 = A_{1} + A_{2} + E_{1}$
in the dihedral group $D_3$
\cite{Atkins}.

For half a flux quantum, certain double degeneracies appear again.
The symmetry assignment in the right gray columns in Fig.\:\ref{fig:spectrum_L3} is based on subtracting $N/2$ quanta from the total momentum $q$. 
This is consistent with the action of the gauge transformation~(\ref{eq:JordanWigner}) on a one-electron wave function with fixed momentum.
The transformed Hamiltonian $H' = Q H Q^{-1}$ becomes real and the Peierls phase at half a flux quantum enforces a ``twisted'' boundary condition on the ring-closing link between sites $s = L$ and $1$: 
the corresponding hopping matrix element has the flipped sign
\cite{Zawadzki2017}.
Its one-particle eigenstates are still given by Eq.\,(\ref{eq:one-electron-spectrum}), but with half-integer momenta $q' = q - \Phi/\Phi_0 = q - \frac{1}{2}$
(see also Appendix~\ref{a:half-integer-q}).
Adopting these shifted momentum labels, one can re-group the irreps $A_1$ and $E_{+1}$ that merge at half a flux quantum, into $E_{1/2}$. 
For the momentum $q' = -3/2 \pmod 3$, we use the irrep label $B$ by analogy to integer momenta.
The same rules can be applied to two electrons at half a flux quantum, Fig.\:\ref{fig:spectrum_L3}\,(right).

Comparing the Zeeman spectra for one and two electrons, 
it is striking that the behavior with the magnetic flux is mirrored. 
This is actually related to electron-hole duality that we now turn to. 
In Appendix~\ref{a:duality-and-hopping}, we demonstrate for odd $L$ the duality relation~(\ref{eq:duality-intertwining}) that zero-field energy levels for one hole align with the one-electron sector with half a flux quantum.

\subsubsection{Particle-Hole Duality}
\label{sec:Particle-Hole Duality}

The duality transformation $\Xi$ is not of geometric origin, but may be implemented as a symmetry even when the Zeeman and Coulomb energies are included.
It exchanges particles and holes and flips occupation numbers according to
$\Xi \,n_s\, \Xi^{-1} = 1 - n_s$.
The Hilbert space sector $N = L/2$ with the same number of electrons and holes, is self-dual and admits $\Xi$ as an automorphism. 
Duality can however also be implemented as a (``intertwining'') symmetry between the Hilbert spaces $\mathcal H_N$ and $\mathcal H_{L-N}$ that share the same dimension according to Eq.\,(\ref{eq:dimension-HN}).
In particular, its action on the Hamiltonian does \emph{not} require attributing negative energies to holes \cite{Zirnbauer2021}.

The Coulomb interaction transforms by duality according to
\begin{align}
V 
\mapsto
\frac12 \sum_{s \ne t}
V_{st}
(1-n_s)(1-n_t)
& =
\frac12 \sum_{s \ne t}
V_{st}
\left(
n_s n_t - n_s - n_t + 1
\right)
\nonumber\\
& = V - \sum_{s \ne t} V_{st} n_s 
+ V_C(L)
    \label{eq:Coulomb-vs-duality}
\end{align}
The first term is the untransformed Coulomb repulsion,
while the last term $V_C(L)$ is a constant,
equal to the interaction energy for a fully occupied network $N = L$, hence the label. (See Appendix~\ref{a:Coulomb-scaling} for details.)
Because $V_{st}$ is symmetric and the ring network is homogeneous, we can introduce $V_r = V_{s,s+r}$ which does not depend on $s$. 
This yields $V_C(L) = \frac{L}{2}\sum_{r = 1}^{L-1}V_r$.
In the middle term in Eq.\,(\ref{eq:Coulomb-vs-duality}), the sum $\sum_{t} V_{st}$ can be understood as the Coulomb energy at $s$ in a fully occupied system, as if $n_t = 1$.
Applying the same symmetry argument, we recognize the previous sum which does not depend on $s$.
The sum over $s$ this time gives the total particle number $N$. 
We finally have
\begin{equation}
    \label{eq:result-duality-Coulomb}
    \Xi \, V \, \Xi^{-1} = V + \left( 1 - \frac{2N}{L} \right) V_C(L)
\end{equation}
As expected, the additive correction vanishes for the self-dual case. 
We emphasize that Eq.\,(\ref{eq:result-duality-Coulomb}) does not hold in networks that have differently coordinated sites (e.g., a ring with a side link/ligand or a star with a central site).

To implement duality for the hopping part of the Hamiltonian~(\ref{eq:SystemHamiltonian}), we follow Ref.~\cite{Zirnbauer2021} and introduce its action on the creation and annihilation operators according to
\begin{align}
\label{eq:DefXi}
\Xi 
\, \big(
\sum_s \alpha_s a_s
\big) \,
\Xi^{-1}
=
\sum_s
\mathrm e^{\mathrm i \varphi_s}
\alpha_s^*
a_s^\dagger \,,
\qquad
\Xi \, a_s^\dagger \, \Xi^{-1}
=
\mathrm e^{\mathrm -i\varphi_s} a_s \,
.
\end{align}
In virtue of the fermionic anti-commutation rules, this is consistent with the action on $n_s$ introduced above. 
But note the conjugation of the complex coefficients $\alpha_s$: $\Xi$ is an anti-linear map.
It therefore not only maps particles onto holes, but also reverses the direction of the magnetic field.
Nevertheless, with a proper choice of the local phases,
${\rm e}^{ {\rm i} \varphi_s} = (-1)^{s}$,
duality becomes a symmetry of the Hamiltonian:
\begin{equation}
    \label{eq:duality-is-symmetry-even-L}
    \text{even $L$}: \qquad
    \Xi \, H_N(\Phi) \, \Xi^{-1} = 
    H_{L-N}(\Phi) + \delta V_C(N)
\end{equation}
where the constant $\delta V_C(N)$ is given in Eq.\,(\ref{eq:result-duality-Coulomb}).
For details, see Appendix~\ref{a:duality}.
We also show there that in the occupation number basis, duality acts with a constant sign in each $N$-sector so that one may take
$\Xi \, |n_1, \ldots, n_L\rangle = |1 - n_1, \ldots, 1 - n_L\rangle$.

A convenient way to check if states are dual to each other, is through the Fourier-transformed annihilation and creation operators $c_q$ [see Eq.\,(\ref{eq:one-electron-spectrum}) and Appendix~\ref{a:DualInMomentum}].
Applying the Fourier transformation to Eq.\,\eqref{eq:DefXi} gives
\begin{equation}
\label{eq:fourierXiAction}
\Xi \,c_q\, \Xi^{-1}
=
c_{q+L/2}^{\dagger},
\end{equation}
which is a shift (modulo $L$) by half the Brillouin zone.

\section{Results}
\label{sec:results}
The flux-dependent many-body spectra discussed in the following illustrate the interplay between magnetic flux, geometric symmetries, Coulomb interaction, and particle-hole duality.
We implement the Hamiltonian~(\ref{eq:SystemHamiltonian}) as a Hermitean matrix on the Hilbert space $\mathcal H_N$ with fixed particle number, taking care of signs due to the anti-commuting fermionic operators (Appendix~\ref{a:OneToMany}).
The energies are plotted relative to the average Coulomb interaction $\overline{V_C(N)}$ (Appendix~\ref{a:Coulomb-scaling}).

At zero magnetic flux, the Hamiltonian inherits the full dihedral symmetry of the underlying lattice geometry. 
The eigenstates can therefore be classified according to irreducible representations of $D_L$ which are well known in molecular physics for $L = 4, 5, 6$.
In Table~\ref{t:irreps-L4-L5-L6}, we list these for a single electron.
The general pattern is the same as for $L = 3$ and follows the Bloch states of Eq.\,(\ref{eq:one-electron-spectrum}).
A general result of this level structure is that the ground state for an odd (even) number of electrons is non-degenerate (a doublet), respectively, the only exception being the fully occupied ring.
--
The symmetry transformations are implemented on the $N$-body subspace according to the scheme outlined in Sec.\:\ref{sec:implement-point-group-symmetry}.
It suffices to check that the generating elements of the dihedral group, $C$ and $\sigma$, commute with $H_N$.

\begin{table}[ht]
\begin{center}
\begin{tabular}{ld{2.2}|ld{2.2}|ld{2.2}}
\multicolumn{2}{c|}{$L = 4$}
&
\multicolumn{2}{c|}{$L = 5$}
&
\multicolumn{2}{c}{$L = 6$}
\\
term & \multicolumn{1}{c|}{$E/t$}
        & term & \multicolumn{1}{c|}{$E/t$} 
                & term & \multicolumn{1}{c}{$E/t$}
\\[0.5ex]
\hline
\rule{0pt}{2.5ex}
&
        &&
            & $b_{1}$ & 2
\\
$b_{1}$ & 2
        & $e_2$ & 1.61
            & $e_{2}$ & 1
\\
$e_{1}$ & 0
        & $e_1$ & -0.61
            & $e_{1}$ & -1
\\
$a_{1}$ & -2
        & $a_1$ & -2
            & $a_{1}$ & -2
\end{tabular}
\end{center}
\caption[]{One-electron energies and symmetry labels for ring networks with $L = 4, 5, 6$ sites, using Eq.\,(\ref{eq:one-electron-spectrum}).
\addtext{(For many-electron states, symmetry labels are written in uppercase.)}
}
\label{t:irreps-L4-L5-L6}
\end{table}

\subsection[Example L = 4]{Example $L = 4$}
\begin{figure}
    \centering
    \raisebox{-0.5\height}{%
        \includegraphics[width=0.43\linewidth]{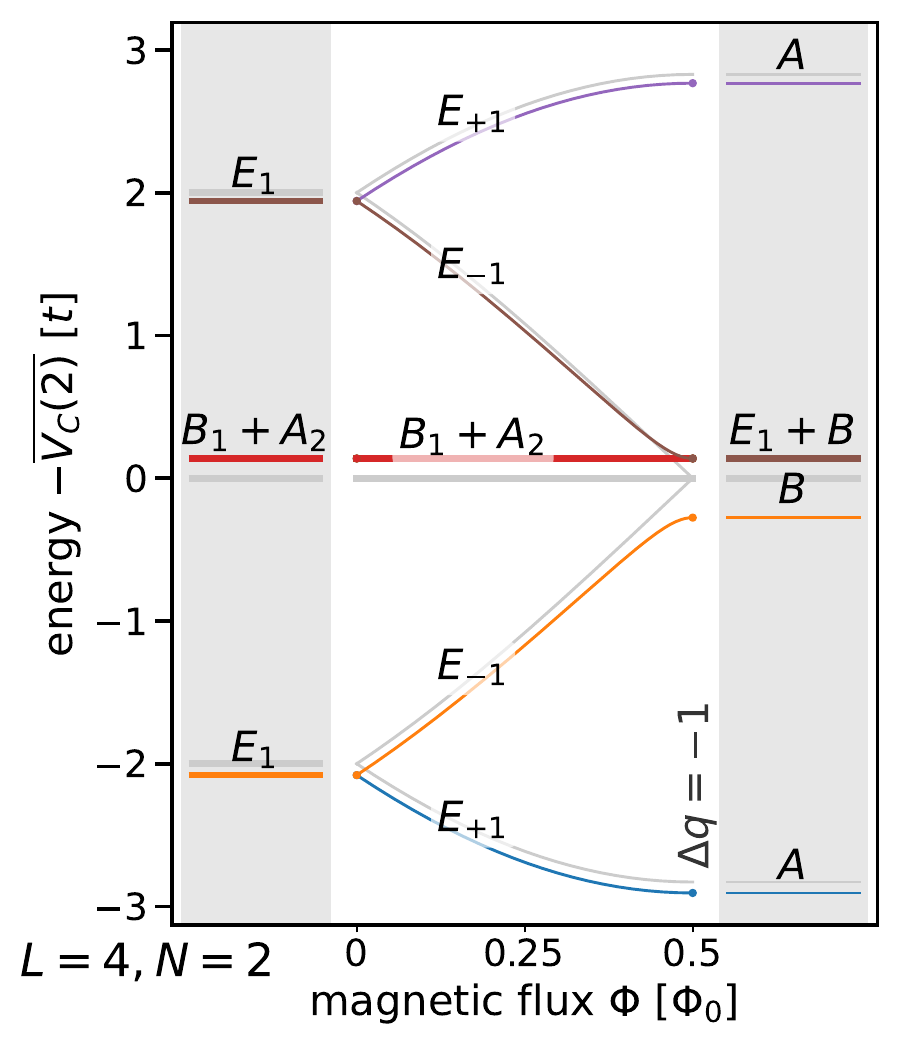}%
    }
    \hspace{0.04\linewidth}
    \raisebox{-0.5\height}{%
        \includegraphics[width=0.35\linewidth]{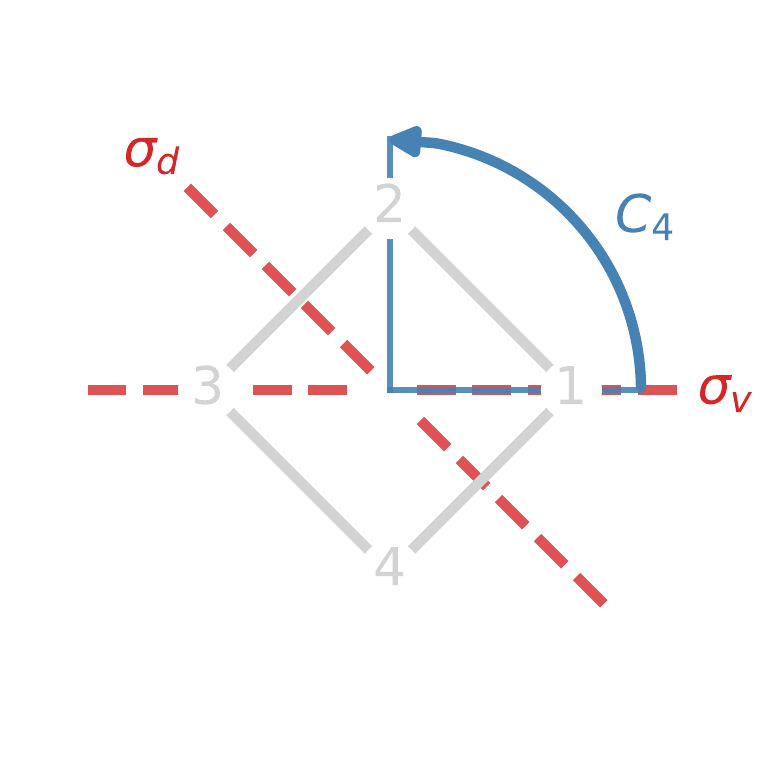}%
    }
    \caption[]{Left: Zeeman spectrum for a square with two electrons. 
    Thick (thin) lines denote two-fold (singly) degenerate levels. 
    We subtract the mean Coulomb interaction $\overline{V_C(2)}$. 
    The irreps $B_{1} + A_{2}$ remain degenerate because they are dual to each other (half-filled network).
    Gray lines indicate energy levels without Coulomb interaction.
    For the symmetry assignment in the right column, a total momentum $-1$ has been subtracted.
    Right: network geometry with its generating symmetry elements.
    For each type $\sigma_v$ and $\sigma_d$, there are two equivalent mirror axes.
    }
    \label{fig:spectrum_L4}
\end{figure}

In Figure~\ref{fig:spectrum_L4}, the flux-dependent spectrum for two electrons on a four-site ring is shown. 
Its overall structure conforms with the \textit{Aufbau} principle: 
the one-electron molecular orbitals $a_{1}$, $e_{1}$, and $b_{1}$ correspond to the momentum states $q=0$, $q=\pm1$, and $q=+2 = -2 \pmod{4}$, respectively (see Tab.\:\ref{t:irreps-L4-L5-L6}).
The ground state would be the configuration $a_{1} e_{1}$ with 
a two-fold degeneracy and symmetry $E_{1}$ if there were no interactions.
The Coulomb interaction mixes this configuration with the upper $E_{1}$ doublet and \emph{lowers} the ground-state energy 
(cf.\ the faded gray lines giving the energies for non-interacting electrons).
Both doublets show a linear Zeeman splitting with persistent currents in opposite directions. 
The magnetic moments can be read off from Fig.\:\ref{fig:E-M-Flow}\,(left) in Appendix~\ref{a:OneToMany}.
For the underlying irreducible representations $E_{+1}$ and $E_{-1}$ \addtext{in nonzero field}, see Appendix~\ref{a:magnetised-dihedral-irrep}.  

The remaining two-electron configurations are $a_{1}b_{1}$ and $e_{1}^2$ and transform according to the one-dimensional irreps $B_{1}$ and $A_{2}$, respectively; 
the Coulomb interaction shifts them up. \deletethis{by the same amount}
The four-site ring provides the simplest example of a self-dual many-body system, where duality $\Xi$ acts within the same Hilbert space and commutes with $H$.
The two states $B_{1}$ and $A_{2}$ are mapped one onto the other by particle-hole exchange. 
This is explained in Appendix~\ref{a:DualInMomentum} using the momentum basis as $B_{1}:\ket{0,2}$ and $A_{2}: \ket{-1,+1}$.
This symmetry protects them from splitting, as manifest in Fig.\:\ref{fig:spectrum_L4}.
Calculating the state $A_{2}$ in the Fock basis with Fourier-transformed operators $c_q^\dag$
\begin{align}
    \label{eq:A2g-in-Fock-basis}
    \ket{-1,+1} 
    &= c_{-1}^\dag c_{+1}^\dag \, \ket{0000}
\nonumber\\
    &= {\textstyle\frac{1}{4}}
    \left( 
    -{\rm i}\, a_1^\dag - a_2^\dag + {\rm i}\, a_3^\dag + a_4^\dag
    \right)
    \left( 
    {\rm i}\, a_1^\dag - a_2^\dag - {\rm i}\, a_3^\dag + a_4^\dag
    \right)
    \ket{0000}
\nonumber\\
    &=
    {\textstyle\frac{{\rm i}}{2}}
    \left(
    \ket{1100}
    + \ket{0110}
    - \ket{1001}
    + \ket{0011}
    \right)
\end{align}
we get the wave function illustrated in Fig.\:\ref{fig:ConfigGraphL4N2} (left).
Note that all its four components have the same Coulomb energy (electrons on neighboring sites), so we deal with an exact eigenstate of $H$.
An alternative proof based on Bloch waves is provided in Appendix~\ref{a:LiftOfDegeneracy}.

Similar to Fig.\:\ref{fig:spectrum_L3}, the symmetry labels in Fig.~\ref{fig:spectrum_L4} (left), right column, illustrate that the classification of the eigenstates at half a flux quantum becomes more ``natural'' when we introduce a shifted total momentum $q' = q - 1$. 
One may view this as a transformation into the frame co-rotating with the persistent current in the ground state.
The symmetry of the latter is assigned from $E_{+1}$ to $A$, while the two states $B_{1} + A_{2}$  become the sector $E_1$.

\addtext{The case $N = 3$ is not displayed because by Eq.\,(\ref{eq:duality-is-symmetry-even-L}), it shows the same spectrum as $N = 1$.}

\subsection[Example L = 5]{Example $L = 5$}
\begin{figure}
\centering
    \raisebox{-0.5\height}{%
\includegraphics[width=0.42\linewidth]{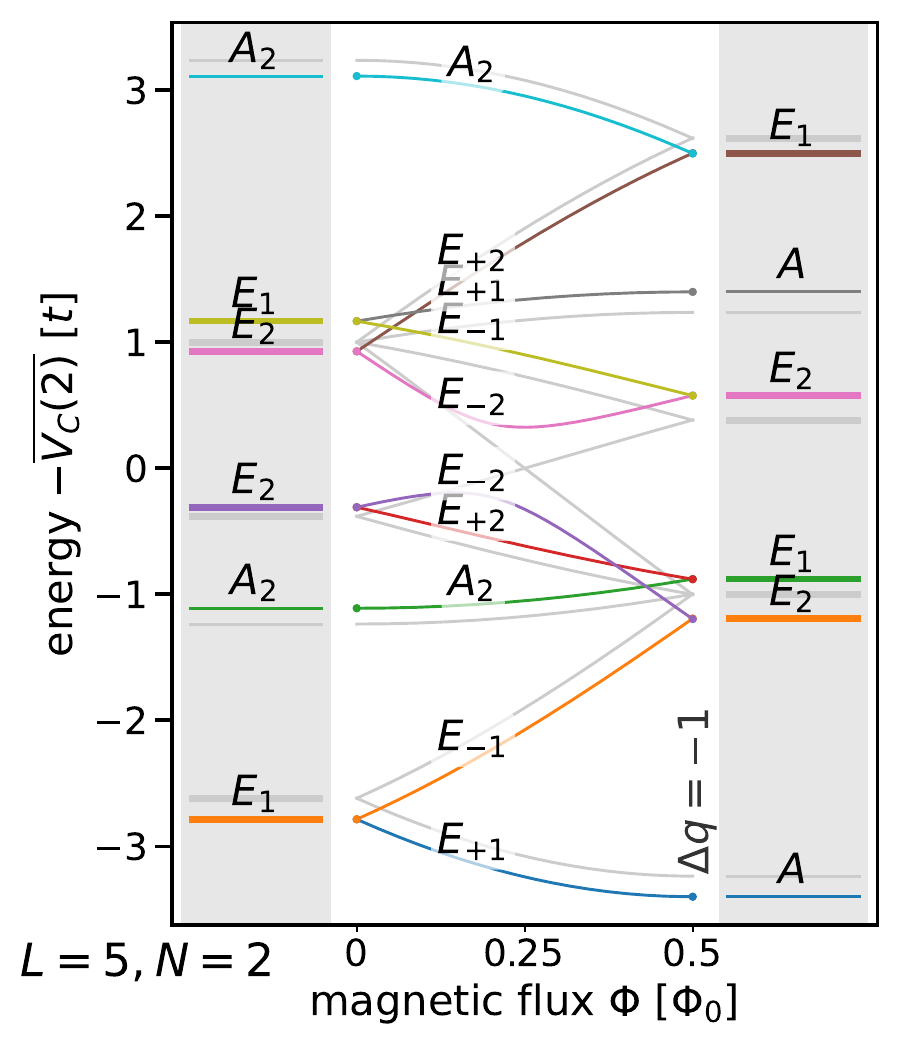}
    }
    \hspace{0.04\linewidth}
    \raisebox{-0.5\height}{%
        \includegraphics[width=0.35\linewidth]{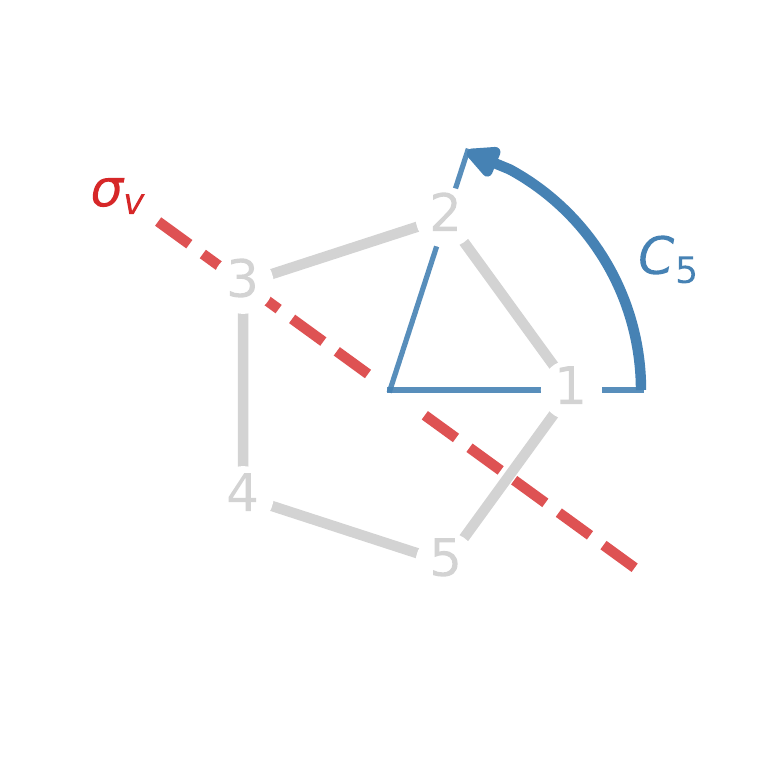}%
    }
    \caption[]{
    Left: Zeeman spectrum for a pentagon ring with two
    electrons. 
    The line thickness indicates the
    degeneracy (single or double). 
    Gray lines indicate the spectra without Coulomb interaction.
    Note the avoided crossing between two $E_{-2}$ terms, while levels with different symmetries cross.
    Right: network geometry and its symmetry elements.
    Five other equivalent $\sigma_v$ axes exist.
    }
    \label{fig:spectrum_L5}
\end{figure}

The Zeeman spectrum of the five-site ring with two electrons is shown in Fig.\:\ref{fig:spectrum_L5}.
Compared to the triangular network, the five-site ring admits two inequivalent particle separations.
For two electrons at
neighboring nodes or separated by one empty site, we have Coulomb energies per pair of $V_1 = V(a) > V_2$, respectively.

In the non-interacting limit, the spectrum exhibits an accidental four-fold degeneracy from the configuration $e_1 e_2$ near $E = t$ (thin gray lines).
This degeneracy is lifted by the Coulomb interaction (in first order) according to the reduction rule $E_1 \times E_2 = E_1 + E_2$ for dihedral irreps in $D_5$. 
It turns out that the $E_1$ term has a larger relative weight for Fock states with electrons on neighboring sites, giving it a positive Coulomb shift.
The Coulomb splitting hence provides information about the spatial correlations encoded in the many-body wave functions. 
For a more detailed discussion see Appendix~\ref{a:LiftOfDegeneracy}.

The Zeeman effect lifts doublet degeneracy,
and as a function of the magnetic flux, terms with different symmetries cross in Fig.\,\ref{fig:spectrum_L5}
(e.g., $A_2$ and $E_{-2}$ near $\Phi = 0.4\,\Phi_0$).
The Coulomb interaction leads to an avoided crossing for the terms $E_{-2}$ near $\Phi = 0.2\,\Phi_0$.
If we plot in a two-dimensional diagram the energy and the magnetization of the states, a dramatic change in magnetic moment appears right at the crossing (see Fig.\:\ref{fig:E-M-Flow} below for $L=5$, $N=2$).
This can also be computed analytically by working out the Coulomb interaction within the two-dimensional subspace spanned by the configurations with momenta $|{-2},0\rangle$ (lower $E_{-2}$) and $\ket{1, 2}$ (upper $E_{-2}$).
The general features of such avoided crossings are discussed in Appendix~\ref{a:avoided-crossings}.

\subsection[Example L = 6]{Example $L = 6$}
\begin{figure}
    \centering
    \includegraphics[width=0.85\linewidth]{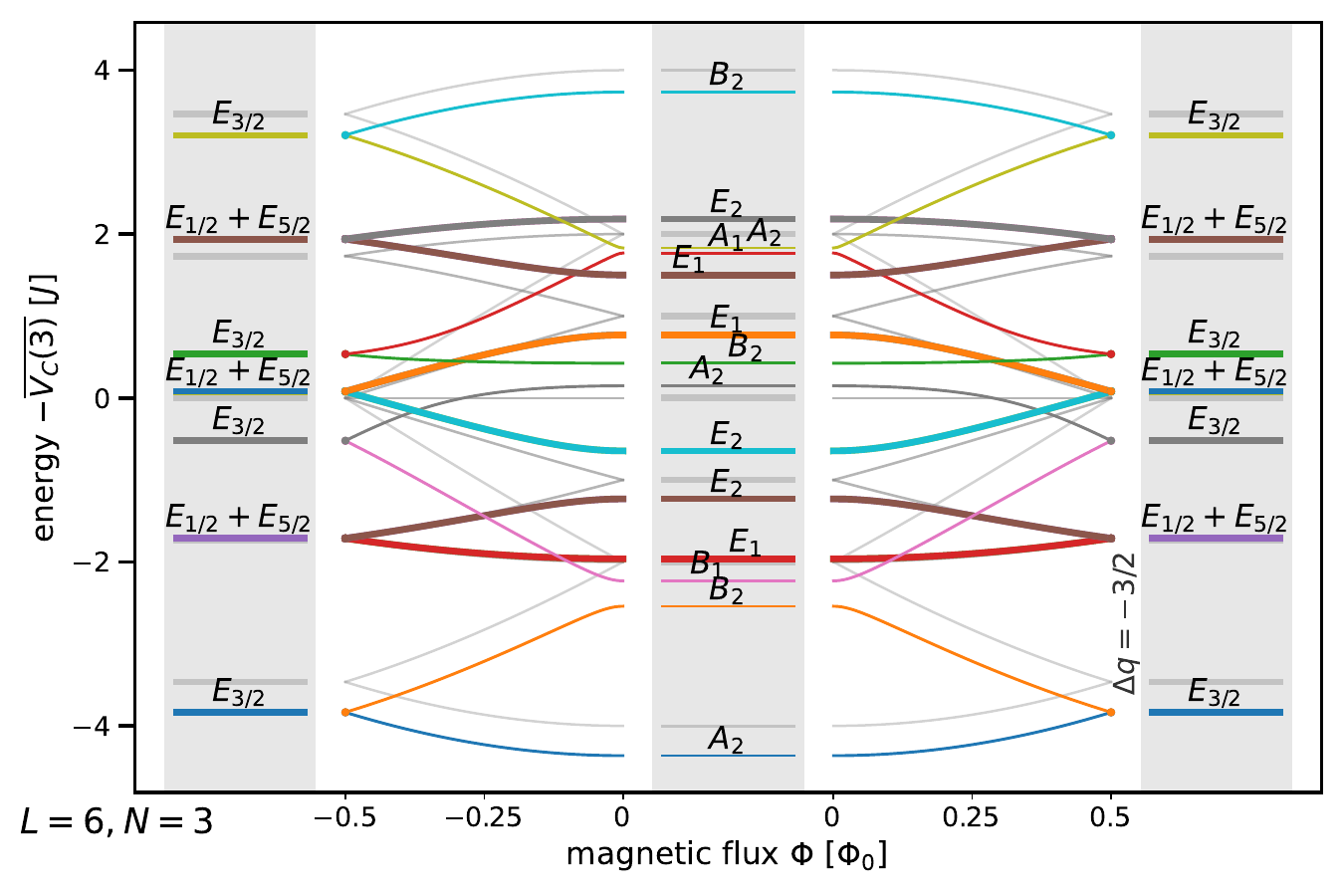}
    \caption[]{%
    Same as Figs.\:\ref{fig:spectrum_L4}, \ref{fig:spectrum_L5} for a hexagonal ring with three
    electrons. 
    Note the doubly degenerate levels (thick lines, $E_1$ and $E_2$): their splitting is forbidden by duality which exchanges the two momentum components.
    }
    \label{fig:spectrum_L6}
\end{figure}

In Fig.\:\ref{fig:spectrum_L6} are shown the Zeeman spectra for the six-site ring with three electrons. 
The detailed behaviour is getting quite involved and is easier to comment on using the energy-magnetisation plane, see Sec.\:\ref{sec:magnetisation-and-rotating-flow} below.
An aspect not encountered so far is that all two-fold degeneracies are not split into opposite momenta, but shift jointly with the magnetic flux. 
This is again due to particle-hole symmetry. 
It turns out that the two momentum eigenstates $\pm q$ in the subspace $E_{q}$ are mapped onto each other by the duality transform $\Xi$. 
One example is discussed in detail in Appendix~\ref{a:DualInMomentum}.

In addition to the geometric symmetries, the Hamiltonian may exhibit additional internal symmetries such as time reversal.
In the absence of magnetic flux, $H_N$ is indeed invariant under time reversal, implemented as element-wise complex conjugation. 
The time-reversal operator $\mathcal T$ therefore acts anti-unitarily and satisfies
\begin{equation}
\mathcal T H_N(\Phi)\mathcal T^{-1} = H_N(-\Phi)
\,.
\end{equation}
This can be seen in Fig.\:\ref{fig:spectrum_L6} where the Zeeman splittings are symmetric under a sign flip of $\Phi$.

\subsection{Illustration of many-Body Configuration Space}
\label{sec:configuration-graph}

\begin{figure}
    \centering
    \includegraphics[width=0.85\textwidth]{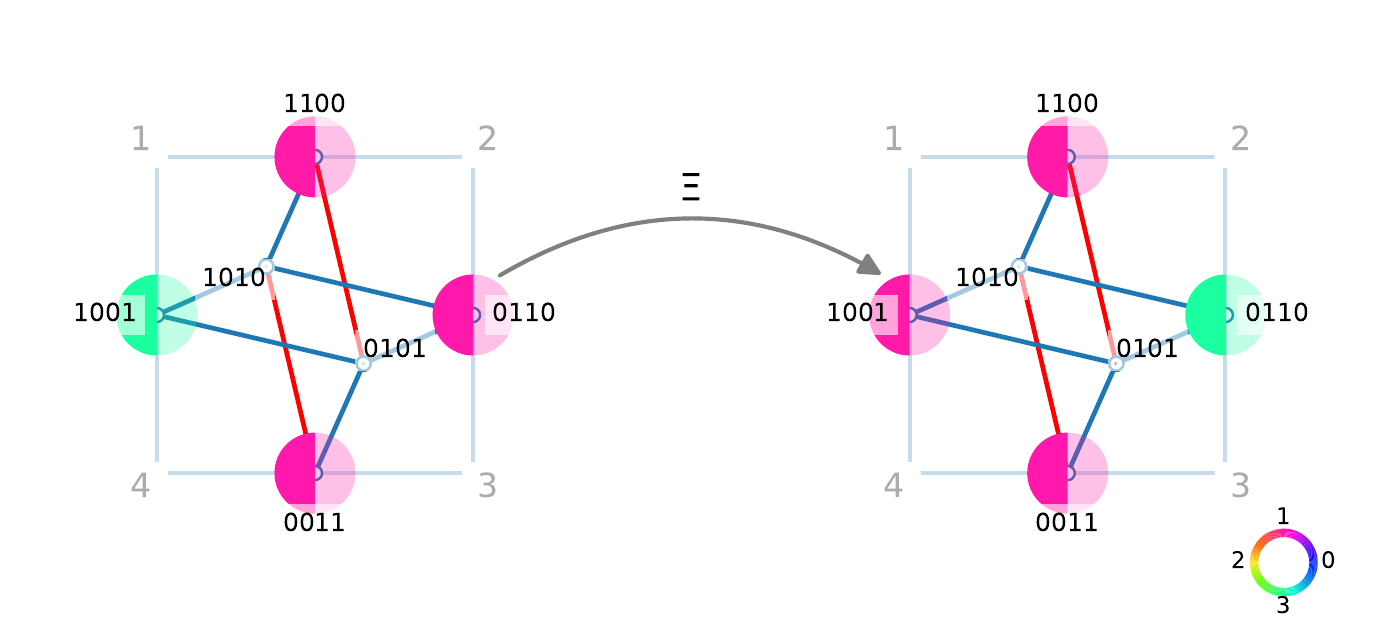}
    \caption{Hopping graphs for $L=4$ and $N=2$. 
    Each node (colored disk) represents a two-particle number state, labeled by the list of site occupation numbers.
    The outer nodes \addtext{$1100$, $0110$, $0011$, \ldots}
    are arranged such that an elementary rotation in the two-particle Hilbert space corresponds to a geometric rotation of the graph. 
    (This is not respected for the inner nodes which have been split spatially to minimize overlapping edges.)
   The edge color indicates the sign of the hopping matrix element between the two states: blue (red) is for $-\hop$ ($+\hop$). 
   The coloured disks visualize the \deletethis{dual} states \addtext{with momentum labels} $\ket{-1,1}$ ($A_{2}$, left, Eq.\,(\ref{eq:A2g-in-Fock-basis})) and $|0,2\rangle$ ($B_{1}$, right). 
   Area and color give probability and phase of the corresponding Fock state, respectively, using the color wheel at bottom right.
   The left half of each disk represents the exact eigenvector with Coulomb interaction, whereas the right half represents the corresponding non-interacting configuration, e.g., $A_{2} = e_1^2$. 
   The two are actually identical because the Coulomb interaction is the same for all occupied \deletethis{nodes} \replaceby{Fock states} (disks with the same area).
   The right panel shows the particle-hole-exchanged wave function obtained by applying the duality transformation $\Xi$ to the state on the left.
   The arrow illustrates the transformation applied to one Fock state.
}
    \label{fig:ConfigGraphL4N2}
\end{figure}

The many-body Hilbert space can be represented by a signed hopping graph $G=(V,E,\eta)$, where each node $v\in V$ corresponds to a Fock state
\begin{equation}
\ket{\mathbf n}
=\ket{n_1,n_2,\dots,n_L},\qquad\sum_s n_s=N .
\end{equation}
Two nodes are connected by an edge $e\in E$ if the corresponding Fock states differ by a single hopping process, see Fig.\:\ref{fig:ConfigGraphL4N2}. 
\addtext{The colors on the edges illustrate}
the edge signature $\eta: E \rightarrow \{-1,+1\}$
determined by the fermionic sign acquired during the hopping process, as explained in Sec.\:\ref{sec:Hubbard-Hamiltonian}.

The particle-hole transformation of Sec.\,\ref{sec:Particle-Hole Duality} acts at half filling within the same Hilbert space and defines a symmetry of the hopping graph with a simple geometric interpretation.
Up to the sign introduced in Eq.\,(\ref{eq:duality-in-occupation-nb-rep}), it maps each Fock state onto its complementary occupation pattern.
Fig.\:\ref{fig:ConfigGraphL4N2} illustrates that this can be implemented as a rotation by $180^\circ$.

The structure of the hopping graph determines whether all edge signs can be removed by a gauge transformation that re-defines the signs of the basis states in $\mathcal H_N$. 
A graph theoretical argument goes as follows: 
Consider a cycle $\gamma$ moving along edges, starting a given node and returning to it.
The length of the cycle (the number of edges) may have the same parity as the product sign $\Pi_{e \in \gamma}\, \eta(e)$. 
If this is true for all cycles, starting from any node, then all edges in the hopping graph can be colored uniformly.
Otherwise, fermionic signs or magnetic phases induce \textit{frustration} on the hopping graph:
its ground state wave function cannot have uniform phases.

The case of two electrons on a ring with $L=5$ sites is such that the hopping graph can be re-colored uniformly. 
\deletethis{Otherwise,}
\replaceby{A two-colorable graph appears}, e.g., for 
$L=4$ and $N=2$, as shown in Fig.\:\ref{fig:ConfigGraphL4N2}.
This implies that the spectrum is symmetric around $E = 0$, unless the underlying Hamiltonian also includes interactions among electrons.

A hopping graph with links of uniform color is pointing to an interesting variant of the molecular network, namely an implementation with bosonic creation and annihilation operators. 
The Fermi sign is absent because the bosonic operators for different sites commute.
One may implement a hard-core repulsion by explicitly forbidding that two (or more) Bosons occupy the same site.
We have checked that the energy spectrum on a 5-site ring with two hard-core Bosons ($\Phi = 0$) is equivalent to the two-Fermion spectrum for half a flux quantum $\Phi = \Phi_0/2$.
This is similar to features seen in Fig.\:\ref{fig:spectrum_L3}, but also holds true if the Coulomb interaction is included.
A more systematic analysis of this Fermion-Boson mapping will be presented elsewhere.

\subsection{Magnetization}
\label{sec:magnetisation-and-rotating-flow}
\begin{figure}
    \centering
    \includegraphics[width=0.45\linewidth]{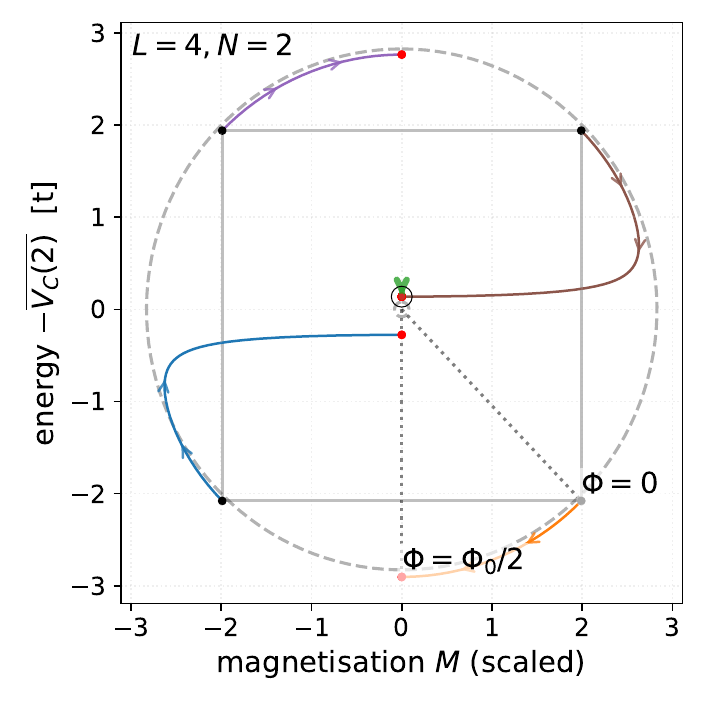}    \includegraphics[width=0.45\linewidth]{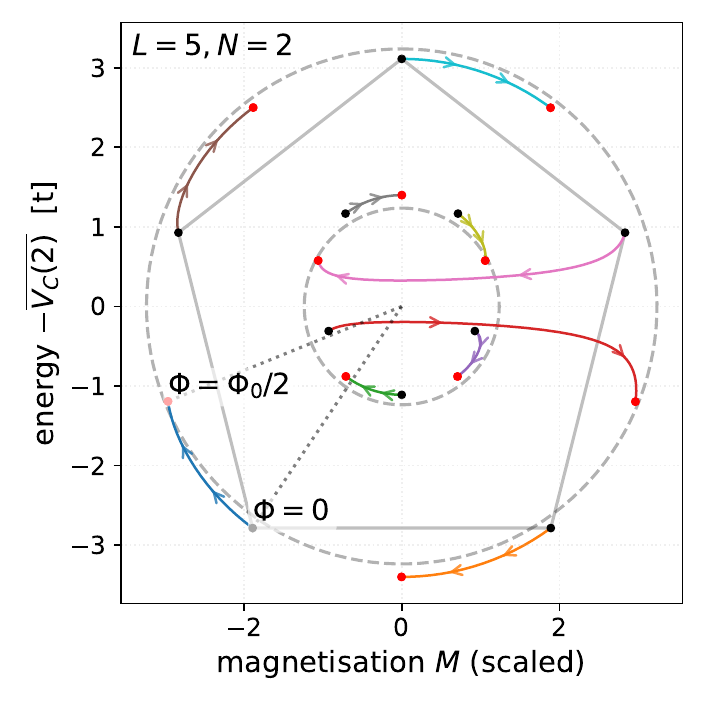}
    \includegraphics[width=0.45\linewidth]{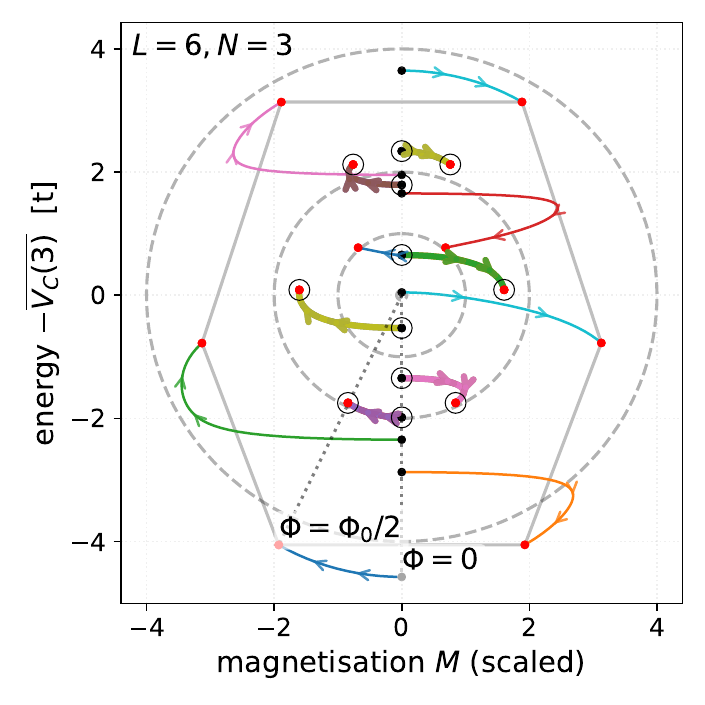}
    \hspace*{8mm}
    \raisebox{+0.05\height}{%
        \includegraphics[width=0.45\linewidth]{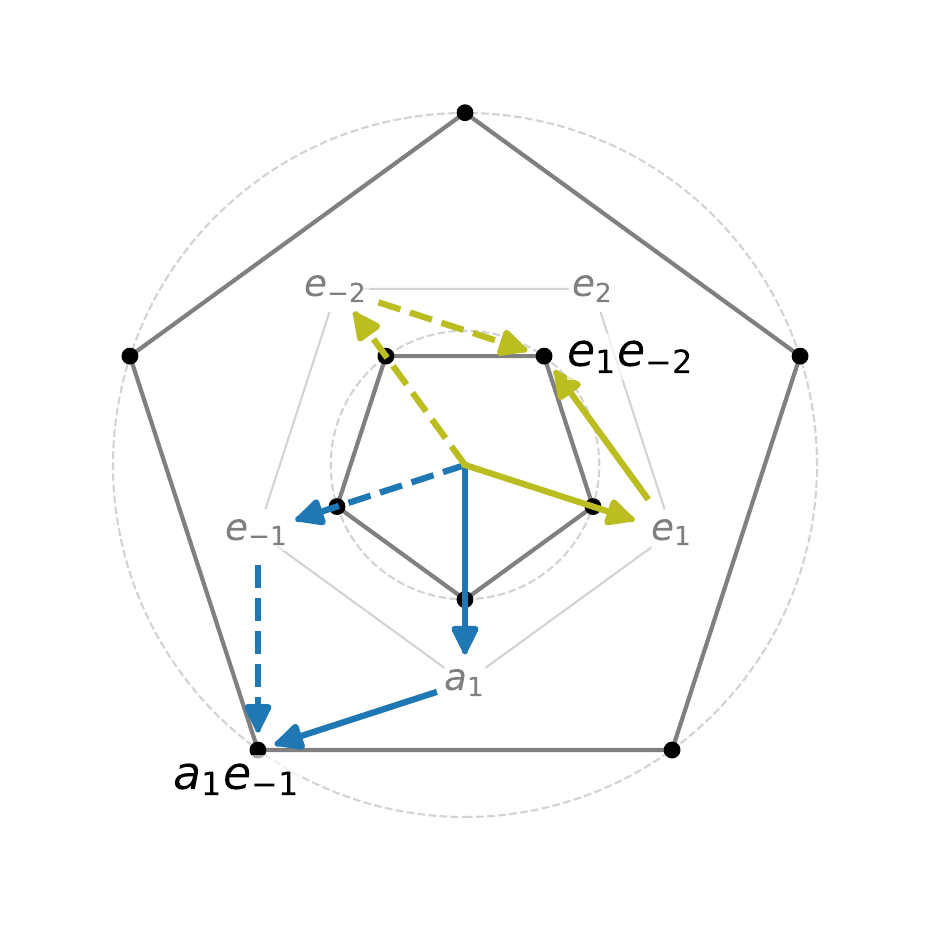}}
    \caption[]{%
    Parametric plot of Zeeman shifts in the energy-magnetisation plane, using the same parameters as Figs.\:\ref{fig:spectrum_L4}--\ref{fig:spectrum_L6}.
    The magnetic flux sweeps $0 \rightarrow \Phi_0/2$ with black (red) colors for initial 
    (final) states, respectively;
    the arrows indicate the movement of the states. 
    The line thickness and encircled points illustrate the degeneracy (single or double).
    In degenerate levels, a basis of eigenstates of the rotation operator $C$ is taken (definite total momentum $q$).
    The polygons that connect states close to the largest circle, would be regular and would rotate rigidly with the flux if the Coulomb interaction were absent.
    States moving along long, nearly horizontal segments indicate avoided crossings.
    The case $L = 4$ (top left) shows a degeneracy near the origin which are the dual states $\ket{A_{2}}$, $\ket{B_{1}}$ that experience no Zeeman and no Coulomb splitting.
    The bottom right figure sketches a systematic scheme for the \textit{Aufbau} principle with 
    two electrons on the five-site ring. 
    The middle pentagon (light gray) locates the one-electron molecular orbitals in the magnetisation-energy plane.
    By adding magnetisation and energy (arrows), one gets two-electron configurations, represented in the same colours as \deletethis{for $L=5, N=2$} \replaceby{in the top right panel.}
    These arrange in two pentagons (dark gray); comparison to the top right scheme illustrate the small shifts arising from the Coulomb interaction.
    }
    \label{fig:E-M-Flow}
\end{figure}
In Fig.\:\ref{fig:E-M-Flow}, the symmetry-adapted magnetisation $M$ is plotted against the energy $E$.
Within each degenerate energy subspace, the operator
$M=-\partial H/\partial\Phi$ is diagonalised to obtain states with a well-defined magnetisation.
The resulting values are rescaled 
such that a one-particle Bloch state would move clockwise around a circle, as predicted by Eqs.\,(\ref{eq:one-electron-spectrum}, \ref{eq:magnetisation-operator}).
The Coulomb interaction moves the points slightly away from these circles.

\paragraph[Example L = 4, N = 2]{Example $L=4,\, N=2$}

For the four-site ring, the many-body spectrum is governed by the momentum sectors $q=0,\pm1,2$. 
At zero magnetic flux, states with $q=\pm1$ are degenerate and form the two-dimensional irreducible representation $E_{1}$. 
These states near energies $E = \pm 2 \,\hop$ carry persistent currents in opposite directions and therefore possess opposite magnetisation, see Eq.\,(\ref{eq:magnetisation-operator}). 
The linear Zeeman splitting emerges in this plot from the tangent vectors close to the dashed circle with opposite vertical components.
A non-degenerate state with $q = 0$ would appear on the axis $M = 0$ and show only a quadratic Zeeman shift like the ground state in Fig.\:\ref{fig:E-M-Flow}\,(bottom left).

Two additional features appear in the central region. The degeneracy at the origin corresponds to the particle–hole conjugate pair $\Xi\ket{A_{2}} = \ket{B_{1}}$, which remains pinned at $M = 0$. 
Furthermore, two $E_{-1}$ branches approach each other coming from opposite values in energy and current; they undergo an avoided crossing. 
The resulting level repulsion is visible in the separation of the blue and brown trajectories which lose very rapidly a large magnetic moment (probably by mixing), as $\Phi \to \Phi_0/2$.

\paragraph[Example L = 5, N = 2]{Example $L=5,\,N=2$}

For the five-site ring, the spectrum is organised by the momentum sectors $q=0,\pm1,\pm2$. 
At zero magnetic flux, there are four states with $q=\pm2$, organised in two degenerate pairs: one pair has parallel one-particle momenta ($\ket{q_1=-2,-1}$ and $\ket{1,2}$, energy $E \approx \hop$) and large magnetization; the other pair's magnetization is smaller ($\ket{-2,0}$ and $\ket{0,2}$, inner circle, $E \approx -0.6\,\hop$).
As the magnetic flux increases, the outer pentagon (and the inner one, not drawn) rotates clockwise.
The states $\ket{1,2}$ (outer right) and $\ket{-2,0}$ (inner left) start to approach in energy and are mixed by the Coulomb interaction, see Appendix~\ref{a:avoided-crossings}). 
Since a crossing in energy is forbidden, a steep change in magnetisation occurs, eventually swapping the states between the outer and inner circles.

\paragraph[Example L = 6, N = 3]{Example $L=6,\,N=3$}

The last example shows the six-site ring with three electrons and features the most complex pattern.
At zero flux, all states have $M=0$ and show quadratic Zeeman shifts.
The states with the lowest and highest energies follow the largest circle. 
Two pairs of states (thin lines starting near $E = 2\,\hop$ and $E = -2.5\,\hop$) rapidly move apart in the magnetic field, indicating an avoided crossing, similar to the feature for $L, N = 4,2$ (top left plot near $E = 0$). 
Indeed, the two lower states with symmetries $B_{1}$ and $B_{2}$ seem to cross over to a linear Zeeman shift (gray lines in Fig.\,\ref{fig:spectrum_L6}), as soon as the field gets strong enough compared to the zero-field Coulomb splitting.
A similar behaviour appears in atomic physics for the Zeeman effect of hyperfine-split levels (Breit-Rabi formula). 
The trajectory of the state that leaves the upper avoided crossing ($E \approx 2\,\hop$) towards a positive magnetization (thin brown) has a slightly different shape and does not end near the large circle.
It is probably undergoing an avoided crossing with the isolated state starting close to the center and ending near $M \approx 3$ (cyan) -- indeed, both may have the $A_{2}$ symmetry.
--
Finally, the doubly degenerate states follow qualitatively similar paths (thick lines) with lesser amplitudes in the magnetization.

The squashed hexagon is drawn here for the states at the final flux value $\Phi_0/2$, and we may assign to its nodes the shifted momenta $q' = \pm 3/2$.
For this flux, the states (marked as red dots) arrange in a pattern with a mirror symmetry $M \leftrightarrow -M$ which could be implemented by complex conjugation. 
(Indeed, the Hamiltonian becomes real after a suitable Jordan-Wigner transformation.)
The top-bottom symmetry $E \leftrightarrow -E$ is broken by the Coulomb interaction, as can be seen by the global downshift of the hexagon with respect to the circle.

\section{Conclusions}

In this paper we have systematically discussed the interplay of discrete symmetries with the Coulomb interaction and a magnetic field in the spectra of molecular networks built from small rings.
The model is based on the Hubbard tight-binding Hamiltonian.
Its representation in a sector with fixed particle number involves Fermi signs for which we have given a geometric intuition in Sec.\:\ref{sec:Hubbard-Hamiltonian}.
The key observables are energy, momentum, and magnetization, the latter we defined following \cite{Meden2003}.
After the introduction of the Jordan-Wigner transformation in Eq.\,\eqref{eq:JordanWigner}, follows the periodicity of the energy spectrum with the magnetic flux.

We discussed the point group of the network, the \textit{Aufbau} principle and the particle-hole conjugation (duality) as a symmetry in the presence of Coulomb and Zeeman interactions.
The point group symmetries provide a natural classification for many-body states via the Mulliken symbols \cite{Mulliken1955} and yield {good} quantum numbers.
\deletethis{with Eq.\,\eqref{eq:good-q-numbers}}
The group theoretical classification explains the origin of spectral degeneracies as well as their lifting under interactions or a magnetic field.
In this context, we have found that half-integer momentum quantum numbers arise \deletethis{at half a flux magnetic quantum $\Phi_0/2$} \replaceby{for half a magnetic flux quantum}.
The field then imposes an antisymmetric (or twisted) rather than periodic boundary condition \replaceby{for the closed ring}.

The flux-dependent spectra presented in Sec.\:\ref{sec:results} demonstrate that the symmetry-based classification provides a practical tool for interpreting interacting many-body states in fine detail. 
It identifies which degeneracies are protected, which crossings are symmetry allowed, and where Coulomb interactions can mix states and produce avoided crossings. 
The particle-hole duality adds an additional symmetry at half-filling and protects selected degeneracies in the spectra, as illustrated in Figs.\:\ref{fig:spectrum_L4} and \ref{fig:spectrum_L6}. 

We have introduced a representation in the energy-magnetisation plane that complements the flux-dependent spectra and visualizes the evolution of many-body states under the Zeeman effect (Fig.\:\ref{fig:E-M-Flow}).
One identifies an intuitive picture for the splitting of degenerate levels, the evolution of persistent currents and the Coulomb-induced state mixing near avoided crossings.

The concepts discussed in this review suggest several directions for future research. 
A natural next step is the extension to larger and more complex molecular networks, including aromatic compounds such as benzene, honeycomb lattices, or random networks. 
Another important feature is to include spin degrees of freedom. 
While spin has been considered in related studies~\cite{Zawadzki2017,Lin2023}, the framework presented here could be extended to include spin-dependent interactions.
Finally, the generalization from isolated molecular rings to open networks by coupling the system to external leads or light fields. 
It would then be interesting to investigate how point-group symmetries and particle-hole duality manifest themselves in electron transport, scattering processes, conductance, and absorption.

\vspace{6pt}

\authorcontributions{%
Carsten Henkel \& Ludwig Schulz authored the text and developed the figures. 
Max Best laid the foundations for the numerical code.}

\funding{%
The research of C.H. is funded by the Deutsche Forschungsgemeinschaft 
(German Research Foundation) within SFB 1636, ID 510943930,
Projects No.\ A01 and A04.
}

\dataavailability{%
No data were generated or used for this study.}

\acknowledgments{%
We thank Anton Bauer, Evgenii Titov, \addtext{and Foudhil Bouakline} for discussion.
}

\conflictsofinterest{}

\appendixtitles{yes} %
\appendixstart
\appendix

\makeatletter
\@addtoreset{equation}{section}
\@addtoreset{figure}{section}
\makeatother
\renewcommand{\theequation}{\thesection\arabic{equation}}
\renewcommand{\thefigure}{\thesection\arabic{figure}}

\section[\appendixname~\thesection]{From One to Many Electrons}
\label{a:OneToMany}

\subsection{Fermionic Operators}
\label{a:a-aDag-Fermi-sign}

The many-electron Hilbert space $\mathcal H_N$ is an antisymmetrized tensor product of single-particle spaces, thus enforcing the Pauli principle.
In this paper, we adopt a formulation entirely in the occupation-number representation using fermionic creation and annihilation operators familiar from second quantisation.
\begin{equation}
    \label{eq:anti-commutators}
\{a_i^{\phantom\dag}, \, a_j^\dagger\} 
= a_i^{\phantom\dag} a_j^\dagger + a_j^\dagger \, a_i^{\phantom\dag}  
= \delta_{ij},
\qquad
\{a_i, \, a_j\} = 0
\,.
\end{equation}
\deletethis{In the present work,} We consider a network consisting of $L$ sites,
each \deletethis{lattice} site $s$ carrying a single localised atomic orbital, and the operator $a_s^\dagger$ creates an electron in it. 
A many-body basis (or Fock) state is specified by the occupation numbers $\{ n_1, n_2, \ldots, n_L \}$
where $n_s\in\{0,1\}$ denotes the occupation of site $s$.
The basis states are generated from the vacuum $|0\rangle = |{\rm vac}\rangle$ by an ordered product of fermionic creation operators,
\begin{equation}
    \label{eq:site-occupation-state}
|n_1,\ldots,n_L\rangle
=
a_1^{\dagger{n_1}}
a_2^{\dagger{n_2}}
\cdots \,
a_L^{\dagger{n_L}}
|0\rangle
\,.
\end{equation}
The fixed ordering of the creation operators uniquely defines the sign convention of the basis states and implies certain signs when hopping processes are implemented.

The states with a single occupied site span the one-particle Hilbert space $\mathcal H_1$ \replaceby{of dimension $L$}.
The tight-binding Hamiltonian restricted to this sector has as its eigenstates the orthonormal molecular orbitals $\phi_q(s)$ labelled by the quantum number $q$. 
Their corresponding creation operators are obtained through a unitary transformation
\begin{equation}
    \label{eq:switch-to-molecular-basis}
c_q^\dagger
= \sum_{s=1}^{L} \phi_q(s)\, a_s^\dagger
\,,
\end{equation}
and also satisfy the anti-commutation rules~(\ref{eq:anti-commutators}). 
We call \deletethis{many-electron} \emph{configurations} the states constructed by occupying more molecular orbitals such that
\begin{equation}
    \label{eq:orbital-configuration}
|q_1, q_2, \ldots, q_N\rangle
=
c_{q_1}^\dagger
c_{q_2}^\dagger
\cdots
c_{q_N}^\dagger
|0\rangle
\,.
\end{equation}
Note again the canonical ordering, arranged in ascending order. 
This is the second-quantised representation of the Slater determinant formed from the occupied molecular orbitals.

As an illustrative example, consider two Fermions on a four-site ring.
The occupation-number basis is given by the ordered lattice-site occupations $\{|1100\rangle,|1010\rangle,\ldots\}$. 
Starting from the Fock state $|1100\rangle$, the hopping process from site $1$ to site $4$ is generated by the operator $a_4^\dagger a_1^{\phantom\dag}$
\begin{equation}
a_4^\dagger a_1^{\phantom\dag} \, |1100\rangle = -|0101\rangle.
\end{equation}
After annihilating the particle on site $1$, the creation operator $a_4^\dagger$ must be permuted past the occupied orbital on site $2$ to restore the canonical ordering of the creation operators in Eq.\,(\ref{eq:site-occupation-state}).
Consequently, the matrix element connecting the states $|1100\rangle$ and $|0101\rangle$ carries a negative sign.

Consider more generally a hopping operation $a_t^\dag a_s^{\phantom\dag}$ applied to the generic Fock state~(\ref{eq:site-occupation-state}).
The annihilator $a_s$ needs to be permuted with the creation operators for labels $1, \ldots, s-1$.
It then combines with $a_s^\dag$ (which must be present) to the unit operator when acting on the vacuum.
Similarly, the creator $a_t^\dag$ anti-commutes with those creation operators with labels $1, \ldots, t-1$, and finally creates a particle at site $t$.
The sign flips compensate for those occupied sites with labels between $1$ and $\min(s-1, t-1)$.
The remaining number is determined by the count $N_{\rm cross}$ of occupied sites with labels between $s$ and $t$, as sketched in Fig.\:\ref{fig:sketches}(right).
Note that ``between'' has to be understood according to the ordered list of sites.
This is to some extent arbitrary and does not have to reflect the actual network connections.
For the ring networks with nearest-neighbor hopping considered here, we have $N_{\rm cross} = 0$, except for the ring-closing link where an electron hops from $s = L$ to $s = 1$ (or back). 
This link gets the additional sign $(-1)^{N-1}$ when the Hubbard Hamiltonian is represented in the subspace with $N$ electrons.

\subsection[\appendixname~\thesubsection]{Coulomb Interaction and Avoided Crossings}
\label{a:avoided-crossings}

In the Hubbard Hamiltonian~(\ref{eq:SystemHamiltonian}), the interaction is diagonal in the occupation number basis~(\ref{eq:site-occupation-state}).
This is no longer true when molecular orbitals~(\ref{eq:orbital-configuration}) or configurations are used.
The interaction between configurations is constrained, however, by selection rules related to the symmetry of the network.
They clarify why crossings appear or are avoided when the magnetic flux is scanned.

The Hamiltonian for a ring network is invariant under the action of the cyclic group $C_L$ when a magnetic flux is present.
It can be block-diagonalized into sectors corresponding to the irreducible representations $\Gamma_i$ of $C_L$.
Let $|\psi_1\rangle \in \Gamma_1$ and $|\psi_2\rangle \in \Gamma_2$. 
For $\Gamma_1\neq\Gamma_2$, symmetry requires the selection rule $\langle\psi_1|H|\psi_2\rangle=0$. 
Consequently, for two states belonging to different symmetry blocks, their energies may cross in the Zeeman spectrum.
(See the levels $E_{-2}, E_{+2}$ and $A_{2}$ near $E = - \hop$ in Fig.\:\ref{fig:spectrum_L5}.)

The situation changes when both states $|\psi_1\rangle, |\psi_2\rangle$ belong to the same irreducible representation.
The coupling matrix element $V(\Phi) = \langle\psi_1|H|\psi_2\rangle$ can be nonzero.
Restricting the Hamiltonian to the subspace spanned by the two states yields the effective Hamiltonian
\begin{equation}
    \label{eq:H_eff_2x2}
    H_{\rm eff}
    =\begin{pmatrix}
        E_1(\Phi) & V(\Phi) \\
        V^*(\Phi) & E_2(\Phi)
    \end{pmatrix},
\end{equation}
where $E_i(\Phi)=\langle\psi_i|H|\psi_i\rangle$. 
Even if the uncoupled levels satisfy $E_1(\Phi) = E_2(\Phi)$ at some flux $\Phi$, the finite coupling prevents an exact crossing, and the level splitting is $\Delta E \ge 2 \,|V(\Phi)|$.
See Fig.\:\ref{fig:spectrum_L5} near $E = 0$ between two levels of type $E_{-2}$.
Since in this case ($L = 5$), there are only two states sharing the same symmetry, the effective Hamiltonian~(\ref{eq:H_eff_2x2}) actually provides an exact calculation.
In the avoided crossing, the eigenstates show a relatively rapid variation with the magnetic flux, as can be checked from their magnetisation shown in Fig.\:\ref{fig:E-M-Flow}.

\subsection[Lift of Degeneracy in L=5, N=2]{Lift of Degeneracy in $L=5$, $N=2$}
\label{a:LiftOfDegeneracy}

We discuss here the splitting of the 4-fold degenerate states near $E = \hop$ in two irreducible representations $E_1$ and $E_2$ (see Fig.\:\ref{fig:spectrum_L5}, left column).
The two-particle wave function provides a qualitative understanding.
For Bloch states
$\phi_q(s)=L^{-1/2} {\rm e}^{2\pi {\rm i}qs/L}$,
two occupied orbitals correspond to the Slater determinant
\begin{equation}
    \Psi_{q_1,q_2}(s,s')
    =\frac{1}{\sqrt2}\left[\phi_{q_1}(s)\phi_{q_2}(s')-\phi_{q_2}(s)\phi_{q_1}(s')\right]
    \,.
\end{equation}
Introducing the center-of-mass coordinate
$S=(s+s')/2$
and the relative coordinate
$d=s-s'$,
the wave function can be rewritten as
\begin{equation}
    \Psi_{q_1,q_2}(S,d)=\frac{2{\rm i}}{\sqrt{2}\, L}e^{2\pi{\rm i}qS/L}\sin (\pi \Delta q \, d/L )
    \,,
\label{eq:relative_wavefunction}
\end{equation}
where
$q = q_1+q_2$
denotes the total and
$\Delta q = \deletethis{q_1-q_2\ }\replaceby{q_2-q_1}$
the relative momentum. 
Since the interaction depends only on the particle separation, the interaction energy is determined entirely by the factor involving the relative coordinate.
The corresponding probability density is
\begin{equation}
\sum_S |\Psi_{q, \Delta q}(S,d)|^2
=\frac{1}{L}\left[1 - \cos(2 \pi \Delta q \, d/L)\right]
\,,
\label{eq:relative_density}
\end{equation}
This vanishes for $d=0$, reflecting the Pauli exclusion principle. 
For a translationally invariant interaction $V(s, s') = V(d)$, the expectation value of the interaction energy becomes
\begin{equation}
    E_{\mathrm{int}}
    =\frac{1}{L}\sum_d V(d)\left[1 - \cos(2 \pi \Delta q \, d/L)\right]
    \,.
\label{eq:interaction_relative}
\end{equation}
which is essentially the discrete Fourier transform of $V(d)$, evaluated at the relative momentum $\Delta q$.

The states $|\replaceby{q_{1}}=1,2\rangle$ and $\ket{-2,-1}$ belong to the irrep $E_2$ and have relative momentum $\Delta q=1$. 
Using Eq.\,(\ref{eq:interaction_relative}), \deletethis{their} \replaceby{they have the} interaction energy \deletethis{is}
\begin{equation}
    E_{\mathrm{int}}(E_2)
    = \frac{1}{L} \sum_d V(d) \left[ 1 - \cos(2\pi d / L) \right]
    \,,
\end{equation}
which must be the same, since the Coulomb interaction commutes with the mirror symmetry that maps the two states one to the other.
Similarly, the states $\ket{-1,2}$ and $\ket{-2,1}$ transform according to the irrep $E_1$ and have relative momentum $\Delta q = 3$. 
The resulting differential interaction energy is
\begin{equation}
    E_{\mathrm{int}}(E_2) - E_{\mathrm{int}}(E_1)
    =\frac{1}{L}\sum_d V(d) \left[
    \cos(6\pi d/L)
    -
    \cos(2\pi d/L) 
    \right]
    \,,
\end{equation}
The only possible distances are $d \in \{1,2\}$, therefore we get 
\begin{align}
    E_{\mathrm{int}}(E_2) - E_{\mathrm{int}}(E_1)
    = \frac{1}{\sqrt5 }[V(2) - V(1)]
    < 0
    \,.
\end{align}
This is actually the exact result, as can be seen in Fig.\:\ref{fig:spectrum_L5} (gray lines) around the energy $E \approx \hop$.
Indeed, since the two irreps differ, the Coulomb potential cannot couple them and only gives diagonal elements (first-order shifts).

For the 4-site ring, the same argument explains why the two-electron states $\ket{q_1=-1,1}$ and $\ket{0,2}$ with irreps $A_{2}$ and $B_{2}$ remain degenerate
(Fig.\:\ref{fig:spectrum_L4}).
They have both $\Delta q = 2$ and thus suffer the same Coulomb shift.

\section[\appendixname~\thesection]{Representations of Dihedral Symmetry}
\label{a:magnetised-dihedral-irrep}
\subsection[\appendixname~\thesubsection]{Zeeman Splitting of $E_q$ Terms}

For a ring with $L$ sites, the cyclic rotation $C$ satisfies
\begin{equation}
C^L = 1 .
    \label{eq:C-power-L}
\end{equation}
Its eigenstates are called momentum states and have quantum numbers
\begin{equation}
C \ket{q} = z_q \ket{q},
\qquad
z_q = {\rm e}^{-2\pi {\rm i} q/L}
\,,
\quad 
q = 0, \pm 1, \ldots  \quad \text{until } |q| \le L/2
\,.
\end{equation}
If only the rotational subgroup $C_L$ is considered, the states $|q\rangle$ and $\ket{-q}$ belong to two distinct one-dimensional irreducible representations, denoted by $E_{+q}$ and $E_{-q}$. 
The case $q = 0$ is denoted $A$, and for even $L$, the labels $q = \pm L/2$ point to the same state with symmetry $B$.

The situation changes when reflection symmetry is included. 
The point group becomes the dihedral $D_L$ and contains a reflection operator $\sigma$ obeying the dihedral relation
\begin{equation}
\sigma C \sigma^{-1} = C^{-1}.
\end{equation}
Applying this relation to a momentum eigenstate gives
\begin{equation}
C(\sigma\ket{q})
= \sigma C^{-1}\ket{q}
= z_q^{-1}\sigma\ket{q}.
\end{equation}
Thus, the reflection maps a state with momentum $q$ onto one with momentum $-q$. 
The two one-dimensional rotational sectors $E_{+q}$ and $E_{-q}$ consequently are paired into a two-dimensional irreducible representation,
\begin{equation}
E_q = E_{+q} \oplus E_{-q}.
\end{equation}
If the reflection symmetry is broken such that rotational symmetry is preserved (e.g., with a magnetic field perpendicular to the ring plane), the representation decomposes into the distinct one-dimensional momentum sectors $E_{+q}$ and $E_{-q}$.

The one-dimensional irreps $A$ and $B$ \deletethis{also} carry \replaceby{an additional} label $1$ ($2$) indicating their even (odd) parity with respect to the mirror plane $\sigma$.
For a ring with an even number of sites, the quantum number for the mirror reflection $\sigma_d$ (going through two opposite edges) is not independent because of the fundamental relation $\sigma_d = C \sigma$ \cite{Tinkham}.
In this case, one also uses in molecular physics a subscript $g$ or $u$ to label the inversion symmetry.
On a ring, this does not provide additional information, however, because the inversion coincides with a $180^\circ$ rotation $C^{L/2}$ with eigenvalue ${\rm e}^{ - \pi{\rm i} q} = (-1)^q$.
--
For the rules to decompose a product representation like $E_1 \times E_1$ into its irreps, see tables in Ref.\:\cite{Atkins}, Resource Section~1.

\subsection[\appendixname~\thesubsection]{Magnetized Representation of full Dihedral Symmetry}
\label{a:half-integer-q}

For a ring threaded by a magnetic flux $\Phi$, the rotational symmetry is represented by a magnetic rotation operator $\widetilde C$. In contrast to the ordinary geometric rotation $C$ of Eq.\,(\ref{eq:C-power-L}), a complete rotation around the ring acquires the Aharonov--Bohm phase
\begin{equation}
\widetilde C^L
=
\mathrm e^{2\pi \mathrm i \,\Phi/\Phi_0}
\,.
\end{equation}
At half a flux quantum, $\Phi=\Phi_0/2$, one obtains
$\widetilde C^L=-1$,
opposite to the ordinary rotation of Eq.\,(\ref{eq:C-power-L}).
This may be called a twisted boundary condition \cite{Zawadzki2017}.
Let $\ket{\psi_q}$ be an eigenstate of $\widetilde C$,
\begin{equation}
\widetilde C \ket{\psi_q}
=
\lambda_q \ket{\psi_q}.
\end{equation}
Then
\begin{equation}
\lambda_q^L=-1.
\end{equation}
For a six-site ring ($L=6$), the solutions are
\begin{equation}
\lambda_q
=
\mathrm e^{-2\pi \mathrm i q / 6},
\qquad
q=\pm\frac12,\,
\pm\frac32,\,
\pm\frac52.
\end{equation}
The representations in the half-flux-quantum case are therefore naturally labelled by half-integer rotational quantum numbers.
In the odd case $L = 5$, the labels $\pm 5/2$ would point to the same state modulo 5, and we may call this 1-dim irrep $B$.
If the rotation $C$ is the only symmetry element, then all irreps are one-dimensional. %

Consider now implementing the mirror reflection $\sigma$ as an anti-linear map $\tilde \sigma = K\, \sigma$ where $K$ is complex conjugation. 
The conjugation undoes the flipping of the magnetic field pseudovector under the geometric reflection,
making $\tilde \sigma$ commute with the Hamiltonian.
We may thus expect to recover the irreps of the full dihedral group $D_L$.
This is not the case, however. 
Despite the intertwining relation
\begin{equation}
	\label{eq:dihedral-intertwining-2}
\tilde\sigma \, C = C^{-1} \, \tilde\sigma
\end{equation}
there exist (complex-valued) states $\ket{q \epsilon}$ ($\epsilon = 1,2$)
which are simultaneous eigenvectors of $C$ and $\tilde\sigma$.
Indeed, according to 
\begin{equation}
\tilde\sigma \, C \ket{q \epsilon} 
= \tilde\sigma \, \lambda_q \ket{q \epsilon} 
= \lambda^*_q (-1)^{\epsilon-1} \ket{q \epsilon} 
\end{equation}
Equation~(\ref{eq:dihedral-intertwining-2}) is satisfied because $\lambda^*_q = \lambda^{-1}_q$ is the eigenvalue of $C^{-1}$.
In this sense, $C$ and $\tilde\sigma$ commute so that we have a faithful, albeit Abelian representation of $D_L$. 
We label its one-dimensional irreps $E_{+q}$ or $E_{-q}$ by the signed momentum;
they join into the 2-dim $E_{q}$ only at zero flux.
(There is actually no point in giving a quantum number like $\epsilon$ above to $\tilde\sigma$ because the eigenvalue changes, for this anti-linear map, with any redefinition of the eigenstate's global phase. 
This is why in Figs.\:\ref{fig:spectrum_L3}--\ref{fig:spectrum_L6}, the subscript for the $\sigma$ symmetry only appears at zero magnetic flux.)

\section{Duality in Fock Space}
\label{a:duality}

\subsection{Intertwining with Hopping Hamiltonian}
\label{a:duality-and-hopping}

A particular phase convention has to be applied in the dual transformation
\begin{equation}
\label{eq:PhaseChoice}
\Xi \, a_s \, \Xi^{-1} = (-1)^{s} \,  a^\dag_s \,.
\end{equation}
in order to have $\Xi$ commute with the hopping part of the Hamiltonian \cite{Zirnbauer2021,Zawadzki2017}.
With this choice, the alternating signs cancel with the anti-commuting fermionic operators:
\begin{align}
\Xi \, \sum_{s} 
\left( 
{\rm e}^{{\rm i}\theta} a^\dag_{s+1} a_s 
+ 
{\rm e}^{-{\rm i}\theta} a^\dag_{s} a_{s+1} 
\right) \, \Xi^{-1}
&= 
\sum_{s<L} 
\left( 
{\rm e}^{-{\rm i}\theta} (-a_{s+1} a^\dag_s) +
{\rm e}^{{\rm i}\theta} (-a_{s} a^\dag_{s+1})
\right) \\ %
&\phantom{=}\ 
{}+ (-1)^{L+1}\mathrm{e} ^{\mathrm{i}\theta}  a_1 a_L^\dag 
+  (-1)^{L+1}\mathrm{e} ^{\mathrm{i}\theta}  a_L a_1^\dag \notag \\
&= 
\sum_{s} 
\left( 
{\rm e}^{-{\rm i}\theta} a^\dag_s a_{s+1}
+ 
{\rm e}^{{\rm i}\theta} a^\dag_{s+1} a_{s}
\right) \quad \text{if $L$ is even}
\end{align}
We have to consider separately the hopping along the ring-closing edge $L\leftrightarrow 1$.
For even $L$, the operators $a_1$ and $a_L$ transform in Eq.\,(\ref{eq:PhaseChoice})
with opposite signs so that the same Hamiltonian emerges even with a magnetic flux.
This is not possible for odd $L$, but can be repaired by mapping to a
Hamiltonian with a different flux. 
Recalling the Jordan-Wigner transformation~(\ref{eq:JordanWigner}),
half a flux quantum gives a real-valued Hamiltonian where the hopping sign at the ring-closing edge is flipped compared to zero flux.
We thus get the ``intertwining relation'' 
\begin{equation}
\label{eq:duality-intertwining}
\text{odd $L$:}\qquad
\Xi \, H_N(\Phi) \, \Xi^{-1} 
\simeq 
H_{L-N}(\tfrac12 \Phi_0 + \Phi) + \delta V_C(N) 
\end{equation}
where the equivalence $\simeq$ is meant up to a unitary transformation,
and $\delta V_C(N)$ is the additive constant in Eq.\,(\ref{eq:result-duality-Coulomb}).
If time reversal is applied on the lhs, one gets $\Xi \, H_N(-\Phi) \, \Xi^{-1}$ with the opposite flux. 
This duality relation is the reason why the Zeeman spectra for $N = 1, 2$ are mirror images, as seen in Fig.\:\ref{fig:spectrum_L3}.

\subsection{Action on Number States}
\label{a:duality-at-fixed-N}

We check here that duality acts in the occupation number representation in the intuitive way by flipping the latter.
For the empty network, we adopt the convention $\Xi\ket{0} = \ket{1,\dots,1}$.
Recalling from Eq.\,(\ref{eq:site-occupation-state}) the construction of a generic basis state by the repeated action of creation operators
\begin{equation}
\label{eq:populate-ket}
\ket{n_{1}, \ldots, n_{L}} = a_1^{\dag n_1} \ldots \, a_L^{\dag n_L} \ket{0}
\end{equation}
We thus get %
\begin{equation}
\Xi\,
\ket{
n_{1},\dots,n_{L}
}
=
(-1)^{n_1} \, a_1^{n_1} \ldots 
(-1)^{s n_s} \, a_s^{n_s} \ldots 
(-1)^{L n_L} a_L^{n_L}
\ket{1,\dots,1}.
    \label{eq:duality-on-Fock-state-0}
\end{equation}
By the same argument as in Appendix\:\ref{a:a-aDag-Fermi-sign}, the operator $a_s$ annihilating the Fermion on site $s$ has to be shifted through $s-1$ creation operators $a^\dag_t$ with $1 \le t < s$, giving a sign $(-1)^{(s-1)n_s}$.
Performing this from right to left with the operators in Eq.\,(\ref{eq:duality-on-Fock-state-0}), 
the duality signs $(-1)^{s n_s}$ are partly compensated, and one gets
\begin{equation}
\Xi\,
\ket{
n_{1},\dots,n_{L}
}
=
(-1)^N
\ket{\bar n_{1}, \ldots, \bar n_{L}}
\,,
\quad 
\bar n_s = 1 - n_s
\label{eq:duality-in-occupation-nb-rep}
\end{equation}
where $N$ is the total particle number.

\subsection{Action in Momentum Space}
\label{a:DualInMomentum}

For a half-filled ring network, the action of the particle-hole transformation on momentum-space operators in Eq.\,(\ref{eq:fourierXiAction}) can be derived straightforwardly. Starting from the Fourier representation of the annihilation operator, one obtains
\begin{align}
\Xi \, c_q \, \Xi^{-1}
&=
\frac{1}{\sqrt{L}}\sum_s \Xi \,{\rm e}^{2\pi{\rm i} qs/L} a_s \, \Xi^{-1}
=
\frac{1}{\sqrt{L}}\sum_s {\rm e}^{-2\pi{\rm i} qs/L}(-1)^s a_s^\dagger
\nonumber\\
&=
\frac{1}{\sqrt{L}}\sum_s {\rm e}^{-2\pi{\rm i} (s/L)(q+L/2))} a_s^\dagger
=
c_{q+L/2}^{\dagger}
\,.
\label{eq-a:fourierXiAction}
\end{align}
using the staggered phase of Eq.\,(\ref{eq:PhaseChoice}).
The duality transform shifts the momentum quantum number by half of the Brillouin zone.
This flips in sign the energy~(\ref{eq:one-electron-spectrum}) of the Bloch orbital and is the key ingredient to make $\Xi$ commute with the hopping Hamiltonian \cite{Zirnbauer2021}.
Furthermore, we define the action of $\Xi$ on the momentum-space vacuum such that the empty state is mapped onto the completely filled Fermi sea,
$\Xi \,|{\rm vac}\rangle = |F\rangle $. 

With this result, the origin of the particle-hole induced degeneracies in Fig.\:\ref{fig:spectrum_L4} becomes transparent. 
Consider the four-site ring, where the allowed orbitals have momentum quantum numbers $0,\pm1,2$. 
One of the degenerate Fock states in the momentum space is
\begin{equation}
|A_{2}\rangle = |q_1 = -1,+1\rangle\,.
\end{equation}
Applying the particle-hole transformation~(\ref{eq-a:fourierXiAction}) yields
\begin{align}
\Xi \, \ket{-1,+1}
=
\Xi \, c_{-1}^{\dagger} c_{1}^{\dagger} \, \Xi^{-1} \,
\Xi \, \ket{0}
=
c_{1}c_{-1} \ket{F}
=
-\ket{0,2} 
=
-\ket{B_{2}}
\,.
\end{align}
Hence, the states $|{-1},+1\rangle$ and $|0,2\rangle$ form a dual conjugate pair and possess the same energy.

A similar example can be constructed for the half-filled six-site ring with $L=6$ and $N=3$, where the momentum quantum numbers are $0,\pm1,\pm2,3$. 
The \deletethis{Fock states} \replaceby{configurations}
$\ket{\addtext{q_1 = }-1,1,2}$ and $\ket{0,1,3}$ are degenerate. 
Duality yields
\begin{align*}
\Xi\, \ket{{-1},1,2}
=
c_{2}c_{-2}c_{-1}\,\ket{F}
=
- \ket{0,1,3}
\,,
\end{align*}
so that the two are mapped onto each other by particle-hole conjugation.
Since $\Xi$ commutes with $H_3(\Phi)$, this dual pair forms the 2-dim irrep $E_2$ and is protected from being split under the Zeeman or Coulomb interactions (see Fig.\:\ref{fig:spectrum_L6}).

\section{Scaling of Coulomb Interaction for Different Networks}
\label{a:Coulomb-scaling}

For a partially filled network with $N \le L$ electrons, we compute here the Coulomb energy averaged over all fixed-$N$ states.
Consider binning electron pairs in a Fock state, with a total number of $\binom{N}{2}$, according to distance $r$,
the smallest value being the distance $a$ between the nearest neighbours. 
This gives for a given state $\mathbf n = \ket{n_{1}, \ldots, n_{L}}$ a list $\{d_r(\mathbf n) | r= a,\dots\}$ of the number of pairs at distance $r$. 
The Coulomb energy of this state is $V_C(\mathbf n) = \sum_r d_r(\mathbf n) V(r)$. 
Summing over all states $\mathbf n$ with fixed $N$ which are $\binom{L}{N}$ in number,
we may expect that the resulting list $D_r = \sum_{\mathbf n} d_r(\mathbf n)$ is representative of the distance statistics of the fully occupied network ($L=N$) 
that we call $p_r$ in the following.
We thus set $D_r = k p_r$ and fix the factor $k$ by requiring that the sum over $r$ yields the total number of pairs 
\begin{equation}
    \sum_r D_r 
    = \binom{L}{N} \binom{N}{2}, \quad 
    \sum_r  p_r  
    = 
    \binom{L}{2}.
\end{equation}
This yields 
\begin{equation}
    D_r 
    = \frac{ \binom{L}{N} \binom{N }{2}
    }{ 
    \binom{L}{2}
    } p_r
\end{equation}
and provides the $N$-electron Coulomb energy, \textit{averaged} over all states, in the form 
\begin{equation}
    \label{eq:averagedCoul}
    \overline{V_C(N)} 
    = \frac{1}{\binom{L}{N} }\sum_r D_r V(r) 
    = \frac{N (N-1)}{L(L-1)} \sum_r p_r V(r) 
    = \frac{N (N-1)}{L(L-1)} V_C(L)
\end{equation}
where the quantity $V_C(L)$ is the Coulomb energy for the filled network.
We read off from Eq.\,(\ref{eq:averagedCoul}) that the average Coulomb repulsion scales, quite intuitively, with the number of electron pairs.

\begin{figure}
    \centering
    \includegraphics[width=0.66\linewidth]{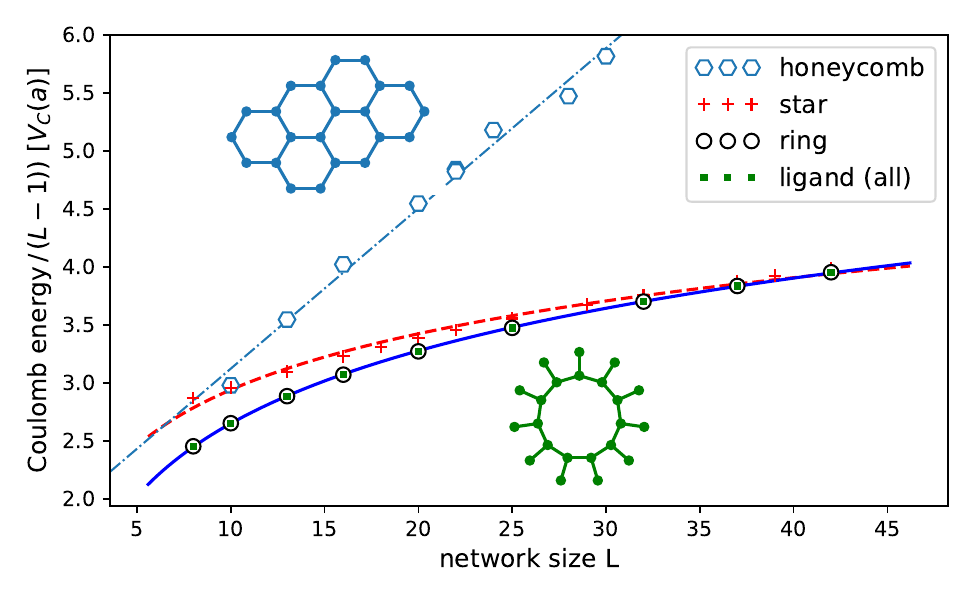}
\caption[]{Scaling of Coulomb energy $V_C(L)$ for a filled network with its size (number of sites $L$), in units of the Coulomb repulsion for two neighbouring electrons (distance $a$) and normalised to the number $L-1$ of ``other'' electrons.
Different shapes beyond the ring are included, as illustrated by the symbols, but all hopping links have the same distance.
The lower sketch gives a 12-ring fully saturated with ligand sites.}
    \label{fig:Coulomb-scaling-with-size}
\end{figure}

In Fig.\:\ref{fig:Coulomb-scaling-with-size}, the network geometry is extended to other shapes like stars or rings with ligands.
An analysis of the Coulomb energy can be done relatively simply in Fock space because it only appears on the diagonal of the Hamiltonian.
We consider in the \deletethis{Fig.\:\ref{fig:Coulomb-scaling-with-size}} \replaceby{Figure} fully occupied networks and the scaling of $V_C(L)$ with the number $L$ of sites.
The scaling in the honeycomb network is different (proportional to $L(L-1)$) which we attribute to the rather planar geometry.
The other shapes share a scaling $(L-1) \log L$ (solid and dashed lines) similar to the ring.
Note that for the star shape, the sites are also arranged on a ring (except for the central site), and that the Coulomb interaction is computed independently of the hopping links.

\reftitle{References}

\isAPAandChicago{}{%

}


\begin{thebibliography}{99}

\bibitem{Spiegelman_2020}
Spiegelman, F.; Tarrat, N.; Cuny, J.; Dontot, L.; Posenitskiy, E.; Martí, C.; Simon, A.; Rapacioli, M.
Density-functional tight-binding: basic concepts and applications to molecules and clusters.
\emph{Adv. Phys. X} \textbf{2020}, \emph{5}, 1710252.
\href{https://doi.org/10.1080/23746149.2019.1710252}{doi:10.1080/23746149.2019.1710252}.

\addtext{%
\bibitem{Wigner1934}
Wigner, E. On the Interaction of Electrons in Metals.
\textit{Phys. Rev.} \textbf{1934}, \textit{46}, 1002--11.
\href{https://doi.org/10.1103/PhysRev.46.1002}{doi:10.1103/PhysRev.46.1002}.
}

\bibitem{Hubbard1963}
Hubbard, J. 
Electron Correlations in Narrow Energy Bands.
\textit{Proc. R. Soc. Lond. A} \textbf{1963}, \textit{276}, 238--257.
\href{https://doi.org/10.1098/rspa.1963.0204}{doi:10.1098/rspa.1963.0204}.


\bibitem{Peierls1933}
Peierls, R. 
Zur Theorie des Diamagnetismus von Leitungselektronen.
\textit{Z. Phys.}  \textbf{1933}, \textit{80}, 763--91.\href{https://DOI.org/10.1007/BF01342591}{DOI:10.1007/BF01342591}.

\addtext{%
\bibitem{Wigner_molecules}
Ziani, N.; Cavaliere, F.; Becerra, K.; Sassetti, M.
A Short Review of One-Dimensional Wigner Crystallization. \textit{Crystals} \textbf{2021} \textit{11}(1), 20.
\href{https://doi.org/10.3390/cryst11010020}{doi:10.3390/cryst11010020}.
}

\bibitem{Lin2023}
Lin, L.; Ke, Y.; Lee, C.
Topological invariants for interacting systems: From twisted boundary conditions to center-of-mass momentum.
\textit{Phys. Rev. B} \textbf{2023}, \textit{107}, 125161.
\href{https://doi.org/10.1103/PhysRevB.107.125161}{doi:10.1103/PhysRevB.107.125161}.

\bibitem{Meden2003}
Meden, V.; Schollw\"ock, U.
Persistent Currents in Mesoscopic Rings: A Numerical and Renormalization Group Study.
\textit{Phys. Rev. B} \textbf{2003}, \textit{67}, 035106.
\href{https://doi.org/10.1103/PhysRevB.67.035106}{doi:10.1103/PhysRevB.67.035106}.

\bibitem{London1937}
London, F. Th\'eorie Quantique des Courants Interatomiques dans les Combinaisons Aromatiques.
\textit{J. Phys. Radium} \textbf{1937}, \textit{8}, 397--409.
\href{https://doi.org/10.1051/jphysrad:01937008010039700}{doi:10.1051/jphysrad:01937008010039700}.

\bibitem{Zirnbauer2021}
Zirnbauer, M.R. Particle--Hole Symmetries in Condensed Matter.
\textit{J. Math. Phys.} \textbf{2021}, \textit{62}, 021101.
\href{https://doi.org/10.1063/5.0035358}{doi:10.1063/5.0035358}.

\bibitem{Mulliken1955}
Mulliken, R.S. Report on Notation for the Spectra of Polyatomic Molecules.
\textit{J. Chem. Phys.} \textbf{1955}, \textit{23}, 1997--2011.
\href{https://doi.org/10.1063/1.1740655}{doi:\linebreak[0]10.1063/\linebreak[0]1.1740655}.

\bibitem{Atkins}
Atkins, P.; Friedman, R.
\textit{Molecular Quantum Mechanics}; Oxford University Press: Oxford, 2010; 5th ed.
\\
\href{https://doi.org/10.1093/hesc/9780199541423.001.0001}{doi:10.1093/hesc/9780199541423.001.0001}.

\bibitem{Tinkham}
Tinkham, M.
\textit{Group Theory and Quantum Mechanics}; McGraw--Hill: New York, 1964.

\bibitem{Zawadzki2017}
Zawadzki, K.; D'Amico, I.; Oliveira, L. N.
Symmetries and Boundary Conditions with a Twist.
\textit{Braz. J. Phys.} \textbf{2017}, \textit{47}, 488--511.
\href{https://doi.org/10.1007/s13538-017-0517-9}{doi:10.1007/s13538-017-0517-9}.

\bibitem{AharonovBohm1959}
Aharonov, Y.; Bohm, D. Significance of Electromagnetic Potentials in the Quantum Theory.
\textit{Phys. Rev.} \textbf{1959}, \textit{115}, 485--491.
\href{https://doi.org/10.1103/PhysRev.115.485}{doi:10.1103/PhysRev.115.485}.



\end{thebibliography}
\end{document}